\documentclass[aoas,preprint]{imsart}

\RequirePackage[OT1]{fontenc}
\RequirePackage{amsthm,amsmath,amsfonts,bm,float,afterpage,bbm,tensor}
\RequirePackage{natbib}
\RequirePackage{adjustbox}
\RequirePackage[colorlinks,citecolor=blue,urlcolor=blue]{hyperref}
\numberwithin{equation}{section}
\usepackage{pdfpages}

\startlocaldefs

\newcommand{\nc}{\newcommand}

\nc{\E}{\mathbb{E}}
\nc{\Var}{\mbox{Var}}
\nc{\Cov}{\mbox{Cov}}
\nc{\R}{\mathbb{R}}
\nc{\pois}{\mbox{Poisson}}
\nc{\binomial}{\mbox{Binomial}}
\nc{\normal}{\mbox{Normal}}
\nc{\multinomial}{\mbox{Multinomial}}

\nc{\yij}{y_{ij}}
\nc{\yik}{y_{ik}}
\nc{\yijk}{y_{ijk}}
\nc{\yi}{\mathbf{y}_i}

\nc{\Y}{\mathbf{Y}}
\nc{\Yh}{\mathbf{Y}_h}

\nc{\bij}{b_{ij}}
\nc{\bi}{\mathbf{b}_i}

\nc{\betaij}{\beta_{ij}}
\nc{\betaj}{\bm{\beta}_j}

\nc{\alphaij}{\alpha_{ij}}
\nc{\alphaj}{\bm{\alpha}_j}

\nc{\xij}{\mathbf{x}_{ij}}

\nc{\mXi}{\mathbf{X}_i}
\nc{\mXk}{\mathbf{X}_k}
\nc{\mXh}{\mathbf{X}_h}
\nc{\mXkh}{\mathbf{X}_{kh}}

\nc{\muij}{\mu_{ij}}
\nc{\mui}{\bm{\mu}_i}
\nc{\nuij}{\nu_{ij}}
\nc{\nui}{\bm{\nu}_i}

\nc{\lambdaij}{\lambda_{ij}}
\nc{\lambdaj}{\lambda_{j}}
\nc{\lambdak}{\lambda_{k}}
\nc{\lambdaik}{\lambda_{ik}}
\nc{\lambdai}{\bm{\lambda}_i}

\nc{\lambdatildeij}{\tilde{\lambda}_{ij}}
\nc{\lambdatildeik}{\tilde{\lambda}_{ik}}
\nc{\lambdatildei}{\bm{\tilde{\lambda}}_i}
\nc{\Lambdai}{\bm{\Lambda}_i}
\nc{\Lambdatildei}{\bm{\tilde{\Lambda}}_i}

\nc{\D}{\mathbf{D}}

\nc{\dij}{d_{ij}}
\nc{\dik}{d_{ik}}
\nc{\dic}{d_{ic}}
\nc{\djk}{d_{jk}}
\nc{\djj}{d_{jj}}
\nc{\dkk}{d_{kk}}
\nc{\dikc}{d_{ikc}}
\nc{\dihkc}{d_{ihkc}}

\nc{\ti}{t_i}
\nc{\tik}{t_{ik}}

\nc{\deltai}{\delta_i}
\nc{\deltaij}{\delta_{ij}}
\nc{\deltaic}{\delta_{ic}}
\nc{\deltaik}{\delta_{ik}}
\nc{\deltaj}{\delta_j}
\nc{\deltaiA}{\delta_{iA}}
\nc{\deltaiB}{\delta_{iB}}

\nc{\LA}{\mathcal{L}^A}
\nc{\LB}{\mathcal{L}^B}
\nc{\LC}{\mathcal{L}^C}
\nc{\LS}{\mathcal{L}^S}
\nc{\Li}{\mathcal{L}_i}
\nc{\Lh}{\mathcal{L}_h}

\nc{\Lik}{\mathcal{L}_{ik}}
\nc{\LiA}{\mathcal{L}_i^A}
\nc{\LiB}{\mathcal{L}_i^B}
\nc{\LiC}{\mathcal{L}_i^C}
\nc{\LiS}{\mathcal{L}_i^S}

\nc{\tLA}{\tilde{\mathcal{L}}^A}
\nc{\tLB}{\tilde{\mathcal{L}}^B}
\nc{\tLC}{\tilde{\mathcal{L}}^C}
\nc{\tLS}{\tilde{\mathcal{L}}^S}
\nc{\tLi}{\tilde{\mathcal{L}}_i}

\nc{\tLiA}{\tilde{\mathcal{L}}_i^A}
\nc{\tLiB}{\tilde{\mathcal{L}}_i^B}
\nc{\tLiC}{\tilde{\mathcal{L}}_i^C}
\nc{\tLiS}{\tilde{\mathcal{L}}_i^S}

\nc{\tLiAone}{\tilde{\mathcal{L}}_i^{A,1}}
\nc{\tLiBone}{\tilde{\mathcal{L}}_i^{B,1}}
\nc{\tLiAtwo}{\tilde{\mathcal{L}}_i^{A,2}}
\nc{\tLiBtwo}{\tilde{\mathcal{L}}_i^{B,2}}

\nc{\tLij}{\tilde{\mathcal{L}}_i^j}

\nc{\barlambda}{\overline{\lambda}}

\nc{\bz}{\bm{z}}

\nc{\nqx}[2]{\tensor*[_{#1}]{q}{_{#2}}}
\nc{\nqxc}[3]{\tensor*[_{#1}]{q}{^{#3}_{#2}}}

\newcommand{\comment}[1]{}

\endlocaldefs

\begin{document}

\begin{frontmatter}
\title{A flexible Bayesian framework to estimate age- and cause-specific child mortality over time from sample registration data\thanksref{T1}}
\runtitle{Estimating child mortality from SRS data}
\thankstext{T1}{Research reported in this publication was supported by the Eunice Kennedy Shriver National Institute Of Child Health \& Human Development of the National Institutes of Health under Award Number R21HD095451 and the National Institute Of Mental Health of the National Institutes of Health under Award Number DP2MH122405. The content is solely the responsibility of the authors and does not necessarily represent the official views of the National Institutes of Health.  This work was also supported in part by the Bill \& Melinda Gates Foundation, Investment ID: OPP1172551}

\begin{aug}
\author{\fnms{Austin E.} \snm{Schumacher}\thanksref{m1}\ead[label=e1]{aeschuma@uw.edu}},
\author{\fnms{Tyler H.} \snm{McCormick}\thanksref{m1}\ead[label=e2]{tylermc@uw.edu}},
\author{\fnms{Jon} \snm{Wakefield}\thanksref{m1}\ead[label=e3]{jonno@uw.edu}},
\author{\fnms{Yue} \snm{Chu}\thanksref{m3}\ead[label=e5]{chu.282@osu.edu}},
\author{\fnms{Jamie} \snm{Perin}\thanksref{m2}\ead[label=e6]{jperin@jhu.edu}},
\author{\fnms{Francisco} \snm{Villavicencio}\thanksref{m2}\ead[label=e7]{jfvillav1@jhu.edu}},
\author{\fnms{Noah} \snm{Simon}\thanksref{m1}\ead[label=e4]{nrsimon@uw.edu}},
\and
\author{\fnms{Li} \snm{Liu}\thanksref{m2}\ead[label=e8]{lliu26@jhu.edu}}

\runauthor{Schumacher et al.}

\affiliation{University of Washington\thanksmark{m1}, Johns Hopkins University\thanksmark{m2} and The Ohio State University\thanksmark{m3}}

\address{A. E. Schumacher\\
Department of Biostatistics, University of Washington\\
Seattle, WA 98195\\
\printead{e1}
}

\address{T. H. McCormick\\
Departments of Statistics and Sociology, University of Washington\\
Seattle, WA 98195\\
\printead{e2}
}

\address{J. Wakefield\\
Departments of Biostatistics and Statistics, University of Washington\\
Seattle, WA 98195\\
\printead{e3}
}

\address{Y. Chu\\
Department of Sociology, The Ohio State University\\
Columbus, OH 43210\\
\printead{e5}
}

\address{J. Perin\\
Department of International Health, Johns Hopkins Bloomberg School of Public Health\\
Baltimore, MD 21205\\
\printead{e6}
}

\address{F. Villavicencio\\
Department of International Health, Johns Hopkins Bloomberg School of Public Health\\
Baltimore, MD 21205\\
\printead*{e7}
}

\address{N. Simon\\
Department of Biostatistics, University of Washington\\
Seattle, WA 98195\\
\printead*{e4}
}

\address{L. Liu\\
Departments of Population, Family and Reproductive Health and International Health\\
Johns Hopkins Bloomberg School of Public Health\\
Baltimore, MD 21205\\
\printead{e8} \\
\hspace{1cm} \\
\hspace{1cm}
}
\end{aug}

\begin{abstract}

In order to implement disease-specific interventions in young age groups, policy makers in low- and middle-income countries require timely and accurate estimates of age- and cause-specific child mortality. High quality data is not available in settings where these interventions are most needed, but there is a push to create sample registration systems that collect detailed mortality information. Current methods that estimate mortality from this data employ multistage frameworks without rigorous statistical justification that separately estimate all-cause and cause-specific mortality and are not sufficiently adaptable to capture important features of the data. We propose a flexible Bayesian modeling framework to estimate age- and cause-specific child mortality from sample registration data. We provide a theoretical justification for the framework, explore its properties via simulation, and use it to estimate mortality trends using data from the Maternal and Child Health Surveillance System in China.

\end{abstract}


\end{frontmatter}

\section{Introduction}
\label{intro}

The \cite{igme2020levels} estimated that 5.2 million children worldwide died before five years of age in 2019. The international community is increasing investment to develop and implement age-targeted, disease-specific interventions and policy \citep{glass2012ending,aponte2009efficacy,penny2016public, o2009burden, keenan2018azithromycin} that require knowing the patterns of child deaths for multiple causes across age and time. The burden of child deaths is heaviest in low and middle-income countries (LMICs) that lack high quality vital registration (VR) systems to register all births and deaths, creating massive uncertainty. The global health community has been pushing for drastic improvements in child health, most notably with the Sustainable Development Goals (SDGs) from the \cite{un2015sdg}. SDG 3 contains age- and cause-specific targets for reducing mortality. Assessing progress toward these goals and identifying areas for improvement require accurate estimation of cause-specific child mortality.

\begin{figure}[ht]
    \centering
    \includegraphics[width=\textwidth]{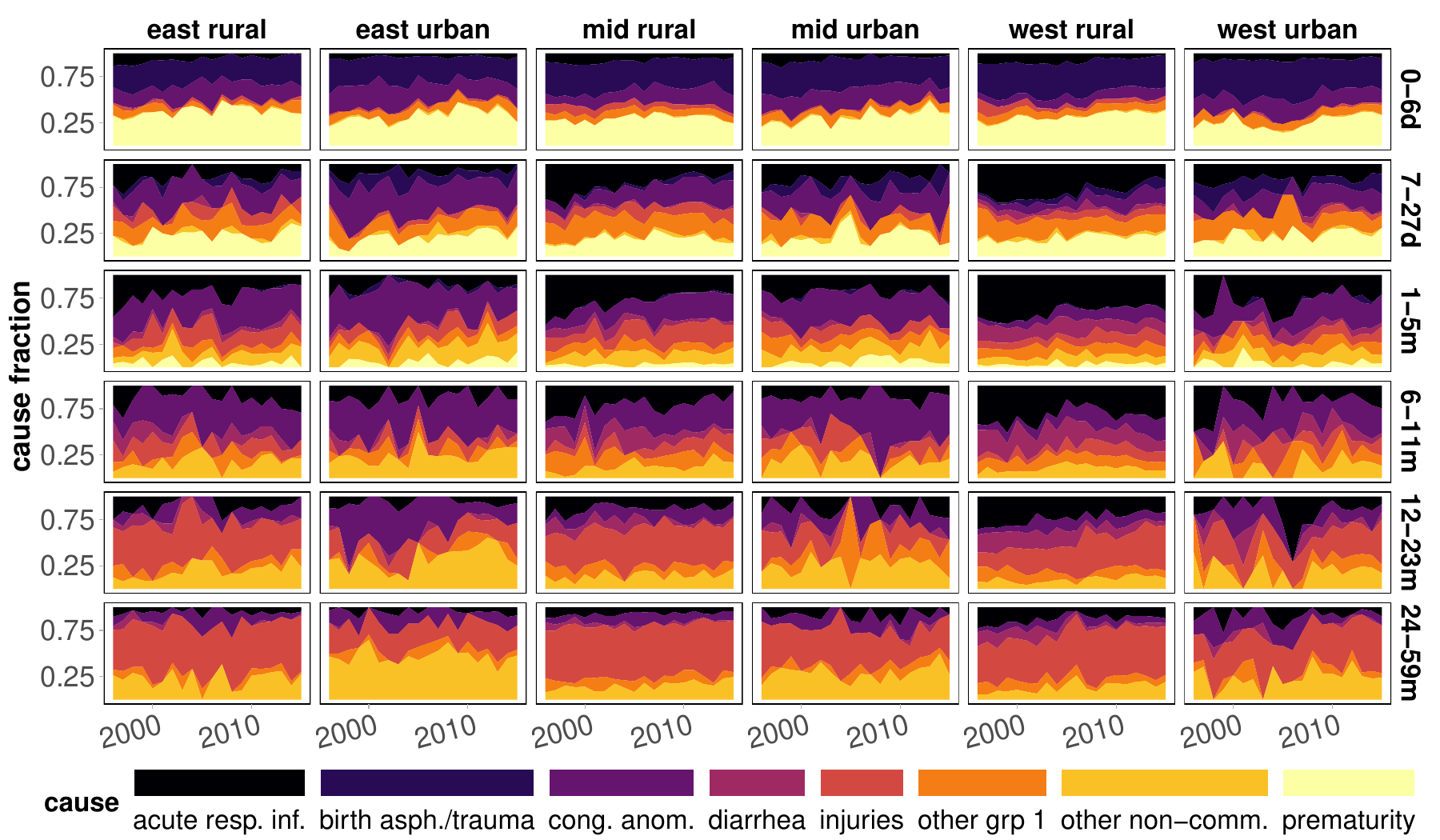}
    \caption{Empirical cause-specific mortality fractions over time by region and age group that were observed in the Maternal and Child Health Surveillance System in China.}
    \label{fig:emp_csmfs}
\end{figure}

High quality VR data is the gold-standard for cause-specific child mortality. In most LMICs, however, VR systems are inadequate \citep{abouzahr2015towards,abouzahr2015civil,mikkelsen2015global,phillips2015well}. Instead, age- and cause-specific mortality data come from sample registration systems (SRS) and national/subnational surveys. Household verbal autopsy (VA) surveys comprise the bulk of national cause-specific mortality data \citep{soleman2006verbal}. However, these data lack continuous monitoring provided by SRS. India and China have led the way in implementing nationally representative SRS \citep{mahapatra2010overview,yang2005mortality,liu2016integrated}, and calls for more and higher quality data collection \citep{bchir2006better,boerma2007health,jha2012counting} have encouraged establishment of SRS in countries such as Indonesia \citep{rao2010mortality} and Mozambique \citep{nkengasong2020improving}. Empirical estimates from SRS are noisy (see Figure \ref{fig:emp_csmfs}), so as SRS data become increasingly available, developing a relevant modeling framework is crucial to produce cause-specific child mortality estimates that provide timely and useful information.

Three main methods are used to estimate cause-specific child mortality that are applicable to SRS data. The first, described by \cite{liu2016global}, models cause-specific mortality fractions (CSMFs) with a multinomial logistic regression model and then multiplies these by all-cause mortality rates estimated in a separate Bayesian framework \citep{you2015global,alkema2014global} to produce cause-specific mortality rates (CSMRs). The second, used in the Global Burden of Disease study and detailed in \cite{vos2020global}, models either rates or probabilities of death separately for each cause with an ad hoc ensemble modeling technique and then combines these with all-cause mortality rates estimated separately using a complex regression model described in \cite{wang2020global}. The third, described by \cite{he2017national}, calculates all-cause mortality rates from a single SRS in China using a 3-year moving average, proportionately scales them so they sum to the all-cause mortality rates estimated in \cite{you2015global}, and then multiplies these by estimated CSMFs at the age-region level that have been smoothed over time using a weighted seven-year moving average. 

The primary issue with these methods is their use of multistage approaches that estimate all-cause and cause-specific mortality in separate, disconnected frameworks that do not ``feed back'' information and uncertainty between stages. A similar issue arises in other fields, for example pharmacokinetic/pharmacodynamic modeling \citep{bennett2001errors} and air pollution modeling \citep{keller2015unified}, which has been addressed by using the full probability model. Joint versus non-joint modeling has been discussed in \cite{plummer2015cuts}. However, the multistage approaches introduced in the previous paragraph leave the models separate. Additionally, they reuse data in both stages, since data sources used for estimating cause-specific mortality are also used in the all-cause mortality models, violating assumed independence between data in the two stages and compromising uncertainty estimates. 

Moreover, these multistage procedures cannot account for some features of the data that are important when modeling cause-specific mortality. For example, correlation between causes can arise when causes share a common underlying factor, such as measles and pertussis---these are both influenced by vaccine access, which may not be included in the model. \cite{vos2020global} and \cite{he2017national} model each cause separately, so no correlation parameters can be included. \cite{liu2016global} do not model correlations, though the framework allows it. Beyond this, the multistage approach in \cite{he2017national} facilitates improperly scaling to the national level using sampling probabilities without accounting for uncertainty. 

Lastly, the approaches of \cite{liu2016global} and \cite{he2017national} produce estimates for broad 0--1 month and 1--59 month age groups that are not sufficient to capture variation in cause of death by age. The cause distribution for the ages within these broad age groups have much heterogeneity (Figure \ref{fig:emp_csmfs}), but these approaches do not disaggregate further because the all-cause mortality estimates come from large global models that only estimate for broad ages, and a lack of smoothing across available variables prevents stable estimation for subgroups with limited data.

To address these issues, our paper brings together multiple strands of research to provide a modeling framework to estimate cause-specific child mortality rates from SRS data. We combine ideas from competing risks, loglinear modeling, and temporal smoothing to construct a framework for developing Bayesian models with an efficient implementation. Our primary contribution is estimating all-cause and cause-specific child mortality in a unified process, rather than an ad hoc multistage approach, that provides more accurate inference and predictions. The flexibility of our framework allows different functional forms to be used as needed depending on context to model mortality at fine granularity by age, cause, region, time, or other strata. We demonstrate building a model using our framework and discuss issues to consider in a motivating example. 

The remainder of this paper is structured as follows. Section \ref{data} describes SRS data in general and details the data from our motivating example, the Maternal and Child Health Surveillance System (MCHSS) in China. Section \ref{theory} develops our proposed framework and makes recommendations for model development. Section \ref{sims} demonstrates improvements over multistage modeling approaches via simulation studies. Section \ref{china} describes the use of our framework to develop and fit a model to the MCHSS data and compares our results to those from \cite{he2017national}. Section \ref{discussion} provides a discussion and future steps.

\section{Data}
\label{data}

Here, we describe the data for our motivating example, the Maternal and Child Health Surveillance System (MCHSS) in China which is China's SRS devoted to maternal and child health. For a description of SRS data in general, see the supplementary material.

The MCHSS has a multistage, stratified, clustered sampling design that is regarded to be representative of the six region-residency strata, henceforth referred to as regions: east urban, east rural, mid urban, mid rural, west urban, and west rural. All children under 5 years of age living in the surveillance sites and all live births of mothers who are either permanent residents of the sites or have lived in the sites for at least 1 year are included. Deaths were assigned a single underlying cause that was ascertained via verbal autopsy, death certificate, or last clinical diagnosis. Deaths were aggregated to the six regions, 20 years of surveillance (1996--2015), six age groups (0--6 days, 7--27 days, 1--5 months, 6--11 months, 12--23 months, and 24--59 months) and 16 cause groups, and then adjusted upwards using a 3 year moving average of under-reporting rates estimated from an annual quality control study. Due to small numbers of deaths, the 16 causes were aggregated into 8 mutually exclusive, collectively exhaustive groups: prematurity, birth asphyxia/trauma, congenital anomalies, other non-communicable, injuries, diarrhea, acute respiratory infections, and other communicable. For prematurity and birth asphyxia/trauma, we deleted as outliers 20 deaths which were in age groups older than 6 months due to implausibility and to prevent unstable estimation from small death counts.

Person-years at risk were not available at the granularity of these six age groups. Therefore, tabulated exposure times were estimated by standard demographic methods described in the supplementary material. Although estimating exposure time is less exact than using recorded person-years under surveillance, exposure times are estimated in many other mortality estimation contexts using demographic methods or by making assumptions, e.g. using mid-year population estimates from the \cite{united2019world}. Furthermore, corrections for complex sampling and underreporting are commonly performed prior to analyses of census, survey, VR, and SRS data along with many other demographic modeling contexts \citep{wang2020global,vos2020global,liu2016global,he2017national,wheldon2013reconstructing}. These and other data issues are discussed in Section \ref{discussion}. For further details on data collection and production of adjusted and aggregated birth and death counts, we refer to the detailed description in \cite{he2017national}. Figure \ref{fig:emp_csmfs} shows empirical CSMFs over time by region and age. We provide death counts, estimated person-years, and log mortality rates by region, age, cause, and time period in the supplementary material.

\section{Statistical framework}
\label{theory}

\subsection{Likelihood}

SRS data arise from a competing risks failure process, as described in \cite{prentice1978analysis}. We assume that each death occurs from a single cause, and the set of causes is mutually exclusive and collectively exhaustive. We will begin by describing the individual-level likelihood and then consider modeling data tabulated by age, time period, and strata. A full derivation is provided in the supplementary material. 

Let $i \in \{1, \dots, n\}$ index individuals and $c \in \{1, \dots, C\}$ index causes of death. Define $T$ as a continuous random variable representing survival time and $J$ as a random variable representing cause of death. We will parameterize survival time by age, i.e. time from birth for each individual. Let $\bz$ be the value of a covariate vector that we assume is fixed for convenience, although this work extends to time-varying covariates in the natural way. We define the cause-specific hazard, which in our case is the mortality rate, as $\lambda_{c}(t| \bz) = \lim_{\Delta t \to 0} P\big(t \leq T < t + \Delta t, J=c | T \geq t, \bz \big) / \Delta t$.

Suppose we have data where $t_i$ is the time of observation, $c_i$ is the cause of death, $\delta_i=1$ if a death is observed and $\delta_i=0$ otherwise (censored), and $\bz_i$ is a vector of fixed covariates for subject $i$. Let $d_{ic}$ indicate that individual $i$ dies from cause $c$. Note that $d_{ic'} = 0$ for all $c' \neq c_i$, and for any censored observations, $d_{ic} = 0 \text{ } \forall \text{ } i, c$. We rewrite the likelihood from \cite{prentice1978analysis}, up to proportionality and with independent censoring, as

\begin{equation}
    \mathcal{L} = \prod_{i=1}^n \prod_{c=1}^C \left[\lambda_{c}(\ti; \bz_i)^{d_{ic}} \exp\left(-\int_0^{t_i}\lambda_c(u; \bz_i)du\right)\right].
\end{equation}

Next, suppose we instead have data tabulated into age groups $k = 1, \dots, K$; here, observation time is parameterized as the age of each individual rather than calendar time. Suppose the data is also tabulated into additional strata $h = 1, \dots, H$. These strata may include time period and region, for example. Notably, an individual may have observation time spent across multiple strata, e.g. observed across different time periods. Using arguments from \cite{holford1976life}, \cite{holford1980analysis}, and \cite{laird1981covariance}, we define $K$ age intervals with breakpoints $0 = \tau_0 < \tau_1 < \dots < \tau_K$. Define $d_{ihkc} = 1$ if individual $i$ dies of cause $c$ in age group $k$ and strata $h$, and define $t_{ihk}$ as the total observation time that individual $i$ is observed in age group $k$ and strata $h$. Note that $d_{ic} = \sum_{k = 1}^K \sum_{h = 1}^H d_{ihkc}$ because a person dies while in only one age group and strata, and additionally an individual's total observation time can be expressed as $\ti = \sum_{k = 1}^K \sum_{h = 1}^H t_{ihk}$. Define the observed data as $y_{hkc} = \sum_{i = 1}^n d_{ihkc}$, the number of deaths in age group $k$ and strata $h$ from cause $c$, and $t_{hk} = \sum_{i = 1}^n t_{ihk}$, the total person-time observed in age group $k$ and strata $h$.

We will assume that cause-specific hazards are constant within each age-strata tabulation group, and for simplicity, we will assume no covariates. Thus, $\lambda_{c}(t_i; \bz_i) = \lambda_{hkc}$ for individual $i$ in strata $h$ with $t_i \in [\tau_{k-1}, \tau_k)$. However, $\lambda_c(t; \bz)$ can depend on covariates and the following derivation extends naturally. The tabulated likelihood is

\begin{equation}
    \mathcal{L} = \prod_{h = 1}^H \prod_{k = 1}^K \prod_{c=1}^C \left[\lambda_{hkc}^{y_{hkc}} \exp\left(- \lambda_{hkc} t_{hk} \right)\right]. \label{eq:fulllike}
\end{equation}

This is identical to the kernel of the likelihood that would arise if $y_{hkc} | \lambda_{hkc}, t_{hk} \sim \pois(\lambda_{hkc} t_{hk})$. Therefore, we can make likelihood-based inference using separate Poisson distributions for each cause and age group.

The product of Poisson likelihoods in (\ref{eq:fulllike}) is equivalent to a model in which the all-cause death counts and person-years have a Poisson distribution and the cause-specific counts conditional on the total death counts have a multinomial distribution. This equivalency of likelihoods is detailed in \cite{lee2017poisson} and is commonly exploited in modeling multinomial count data. This gives rise to a specific multistage modeling specification: First estimate all-cause mortality rates using a Poisson distribution, and then estimate CSMFs conditional on the all-cause death counts using a multinomial distribution. We show the consistency of this two-stage model via simulation in the supplementary material. However, none of the multistage modeling approaches detailed in Section \ref{intro} specify this consistent two-stage likelihood.

Statistical modeling based on our framework provides the flexibility to choose a model that is driven by the study-specific context. It is natural to work with loglinear hazard models due to the constraint that mortality rates must be positive. Then, given a vector of parameters $\bm\eta$, we can construct a model as $\log(\lambda_{hkc}) = f_{hkc}(\bm\eta)$.

\subsection{Model development}
\label{model_devel}

Using our proposed modeling framework, one can specify functional forms for $f_{hkc}(\bm\eta)$ that contain fixed and random effects \citep{breslow1993approximate}, use copula functions \citep{smith2012estimation}, or many other methods. This flexibility is critical to account for the main drivers of mortality, but it also begets the need for careful model construction. We will discuss choosing a functional form for $f_{hkc}(\bm\eta)$ by considering cause of death, time period, age, geography, and the interactions between these variables. We will also discuss accounting for overdispersion as well as model validation and comparison.

\subsubsection{Cause of death}

The relative distribution of causes of death can vary immensely depending on the context of data collection \citep{clark2019verbal}. In order for the competing risks framework to hold, we must have an exhaustive list of mutually exclusive cause groups. For modeling, we need a sufficiently large number of deaths in each group to provide stable estimation. In most situations, we expect that only a small number of cause groups would be plausible to model, and these groups will have distinct differences in mortality. Thus, we recommend using a fixed intercept for each group. 

Another consideration is that causes may be correlated due to covariates that were not collected but influence multiple causes of death, such as environmental factors. Modeling correlations may improve estimates, especially for time periods/regions/ages with little or no data. Unfortunately, estimating correlation parameters is difficult and requires large amounts of data (we discuss this further in Section \ref{discussion}). If data permits stable estimation of correlation parameters, we recommend a hierarchical modeling approach using multivariate normal cause-specific intercepts, i.e. for $C$ causes, $f_{hkc}(\bm\eta)$ would include a vector of fixed effects $\bm{\beta} \sim \mbox{Normal}_C(0, \bm\Sigma)$, where the off-diagonal entries of the covariance matrix $\bm\Sigma$ parameterize the correlations between causes, which can be estimated from the data. A simple example of this model for two cause groups is presented in a simulation in Section \ref{sim2}.

\subsubsection{Time period}
\label{mod_time}

Prompt and accurate time trend estimates support policy enactment, intervention targeting, and resource allocation in an agile fashion \citep{friberg2010sub}. In addition, they allow evaluation of performance toward child survival targets, such as the SDGs, and also serve as important quality indicators for global health statistics and their estimation \citep{walker2007interpreting}. For data with a smaller time span than the MCHSS, modeling can be done via linear trends. As the number of available time points grows, we recommend more flexible methods such as those commonly used in other child mortality models. Second-order random walks are popular \citep{li2019changes}, in which case $f_{hkc}(\bm\eta)$ would include a vector of random effects $(\gamma_1, \dots, \gamma_T)$ for $T$ time points that parameterizes the second differences in time as arising from a normal distribution, i.e. $\gamma_t - 2 \gamma_{t+1} + \gamma_{t+2} \sim \mbox{Normal}(0, \sigma_{\gamma}^2)$, with $\sigma_{\gamma}^2$ parameterizing how quickly the time trend can vary from year to year \citep{rue2005gaussian}. Other options include cubic spline or B-spline models \citep{alkema2014global}; notably, a certain class of cubic splines are equivalent to second-order random walks \citep{speckman2003fully}. Another option is random effects with an autoregressive distribution \citep{chi1989models}. 

\subsubsection{Age}

Mortality trends vary drastically by age; for example, the neonatal period has much higher mortality than the rest of the under-five age range. Age is commonly tabulated into groups, either for data reporting standards, for convenience of modeling, or because of interval censoring. In the past, data for under-5 mortality has been disaggregated into the first year and the combined remaining four years; thankfully, recent data collection and estimation has favored further disaggregation into early- and late-neonatal, along with further breakdown of the 1--4 year period \citep{liu2016global,vos2020global}. Finer age groups allow for more useful estimates to direct health interventions. Treating age as a categorical variable allows flexible, non-monotonic relationships, a modeling choice that is facilitated by the commonly-available tabulated form of SRS data. We recommend using fixed intercepts to capture the main effects of age due to large differences in mortality among commonly available tabulated groups.

\subsubsection{Geography}

In the SDG era, health policy and program decision making are becoming decentralized with many decisions now happening at the district level. Subnational mortality estimates help adapt the development of health statistics to meet changing needs that vary by region \citep{boerma2013public}. Often the geographic information consists of the administrative area in which the child resides. When there are few areas, spatial modeling is difficult because of the paucity of information. In this situation, fixed effects for areas can be used.

If more detailed location information is available, many methods can be used with our framework. Data availability and the scale for reporting estimates should drive this modeling choice. For areal data, popular choices include the reparameterized Besag, York and Molli\'e (BYM) model \citep{riebler2016intuitive} and DAGAR models \citep{datta2019spatial}, amongst many others \citep{heaton2019case}. For point-referenced data, such as individual- or cluster-level data, one can use the Gaussian Markov random field representation of the stochastic partial differential equations approach with a Mat\'ern family for point-referenced data \citep{lindgren2011explicit} or nearest neighbor Gaussian process models \citep{datta2016nonseparable}.

\subsubsection{Interactions}
\label{sub_interactions}

Interactions between cause of death, time period, age, and geography must be considered as we expect differential effects for different combinations of these variables. The amount of data available restricts the number of interactions that can be modeled, which means the context of the data analysis is crucial when choosing a model.

Due to the importance of time trends for global health policy and interventions, modeling different time trends for each age, region, and cause is paramount. For example, mortality rates from injuries are likely to change differently for infants compared to older children, and these may further be different in rural locations compared to urban locations. Modeling these interactions involves allowing for time trends to vary by age-region-cause strata. How to achieve this goal depends on the choice of how main effects are modeled. For example, if temporal trends are modeled via a random walk, one can use different random walk parameters for different combinations of age, region, and cause of death. More complex space-time models can also be used, such as those described in \cite{wakefield2019estimating}.

Non-temporal interactions to consider include the different distributions of causes of death among age groups and regions, as well as different age effects among regions \citep{who2001effect,walker2013global,abdullah2007patterns,snow1997relation}. Accurate modeling of these allows interventions to be targeted to populations in most need. Again, the decision on how to model these interactions depends on how the main effects were chosen to be modeled. For example, if cause of death, age, and geography were all modeled with fixed effects, then one can simply include two-way (and possibly three-way) interactions of these variables. Models with more complicated main effects will necessitate more complex interaction models.

While child mortality is likely to vary by all possible interactions among cause of death, time, age, and geography, one cannot typically include all of these interactions in a model due to the limited amount of available data. Therefore, it is crucial to carefully choose which interactions are most important to model. We believe the end scientific goal should drive the interactions on which to focus. Once the main goals are decided upon, initial data exploration and simulations can be used to determine the extent to which the amount of data allows for different interaction models. We present examples of these exercises in Section \ref{china} with additional details in the supplementary material. 

\subsubsection{Overdispersion}
\label{sub_overdispersion}

Overdispersion is common in child mortality data due to within-strata variability, such as from unobserved covariates, measurement error, non-systematic errors from the data preprocessing steps, and cause misattribution. To account for this, we recommend including observation-specific random effects in $f_{hkc}(\bm\eta)$, i.e. $\epsilon_{hkc} \underset{\text{iid}}{\sim} \mbox{Normal}(0, \sigma_{\epsilon}^2)$. This popular method allows for additional variability, although it does not facilitate differentiating between different sources of overdispersion.

\subsubsection{Model validation and comparison}

Due to the flexibility of our framework and the many decisions that must be made when choosing how to model cause, time, age, geography, and their interactions, models must be checked for adequacy. We recommend plotting standardized residuals grouped by all combinations of each of the strata used in modeling (including two-way, three-way, and higher combinations if necessary) and examining these plots for any distinguishable patterns. Furthermore, one can perform hold out experiments to evaluate predictive performance for important aspects of the model. For example, one can hold out the final year of observations, fit the model, and compare predictions to the held out data in order to evaluate predictive validity of short term time trends. Finally, we recommend comparing the performance of a suite of candidate models via traditional model comparison metrics such as the deviance information criteria (DIC), Watanabe-Akaike information criterion (WAIC), and conditional predictive ordinates (CPO). For a detailed discussion of model comparison, see \cite{gelman2014understanding} and \cite{held2010posterior}.

\section{Simulations}
\label{sims}

This section compares models fit using our framework, which we call unified models, to models fit with multistage frameworks that combine separate cause-specific models. We fit models in this section with the \texttt{INLA} package for fast estimation using integrated nested Laplace approximation \citep{rue2009approximate} in the \texttt{R} statistical computing environment \citep{team2013r}. Models use the default prior distributions in \texttt{INLA}. Full descriptions of each simulation are provided in the supplementary material. Replication code is available at \href{http://www.github.com/aeschuma/SRS-child-mortality}{http://www.github.com/aeschuma/SRS-child-mortality}. 

\subsection{Scenario 1: Extra-Poisson Variability}
\label{sim1}

Let $h \in \{1, \dots, H = 720\}$ index strata with 6 age groups, 6 regions, and 20 years, and let $c \in \{1, \dots, C\}$ index cause. Define $N_h$ as the total exposure time and $y_{hc}$ as the death counts from cause $c$ in strata $h$. Define $\lambda_{hc}$ as the CSMRs, $y_{h+} = \sum_c y_{hc}$ as the all-cause death counts and $\lambda_{h+} = \sum_c \lambda_{hc}$ as the all-cause mortality rates. For $C=8$ causes, we specify 

\begin{align*}
    y_{hc} | \lambda_{hc} &\sim \text{Poisson}\left(N_h \lambda_{hc}\right) \\
    \log(\lambda_{hc}) &= \alpha + \sum_{c' = 2}^C\beta_{c'} \mathbbm{1}_{[c' = c]} + \epsilon_{hc} \\
    \epsilon_{hc}|\sigma_{\epsilon}^2 &\underset{\text{iid}}{\sim} \text{N}(0,\sigma_{\epsilon}^2).
\end{align*}

This exemplifies a scenario with different mortality for each cause along with extra-Poisson variability. We set $\alpha = -5$ and $\beta_c = 0.5$ for all $c$, which yields death counts in the same range as the MCHSS data with a similar overall mean mortality rate of 0.01. We set $\sigma_{\epsilon}^2 = 0.2$, which is approximately equal to the level estimated from our model fit to the MCHSS data described in Section \ref{china}. 

For the multistage model, we estimate all-cause mortality rates using a Poisson GLMM with an overall intercept and IID Normal random effects on strata such that 

\begin{align*}
    y_{h+}| N_h, \lambda_{h+} &\sim \mbox{Poisson}(N_h \lambda_{h+})\\
    \log(\lambda_{h+}) &= \alpha + \gamma_h \\
    \gamma_h|\sigma_{\gamma}^2 &\underset{\text{iid}}{\sim} \mbox{Normal}(0,\sigma_{\gamma}^2).
\end{align*}

To estimate the CSMFs, we use separate Poisson generalized linear mixed models for each cause. These models each have an overall intercept and IID Normal random effects on strata. For each cause $c$, we have 

\begin{align*}
    y_{hc} | N_h, \lambda_{hc} &\sim \mbox{Poisson}(N_h \lambda_{hc})\\
    \log(\lambda_{hc}) &= \alpha_{c} + \xi_{hc}\\
    \xi_{hc} &\underset{\text{iid}}{\sim} \mbox{Normal}(0,\sigma_{\xi c}^2).
\end{align*}

Taking samples $s$ = $1, \dots, 1000$ from the posteriors, for each sample we calculate CSMFs as $\widehat{p}_{hc}^{(s)} = \widehat\lambda_{hc}^{(s)} / \sum_c \widehat\lambda_{hc}^{(s)}$, all-cause mortality rates as $\widehat\lambda_{h+}^{(s)}$, and combine these to calculate log CSMRs as $\log\left(\widehat\lambda_{hc}^{(s)}\right) = \log\left(\widehat\lambda_{h+}^{(s)} \widehat{p}_{hc}^{(s)}\right)$. 

To compare with the multistage model, we fit a unified model correctly specifying the data generating mechanism and draw 1000 posterior samples for each log CSMR. For both the multistage and unified models, we perform 100 simulations each for $N_h \in \{ 1000, 10000, 100000 \}$. These are in the range of the 5th, 50th, and 95th percentiles of exposure time in the MCHSS data, which are $292$, $8749$, and $110974$, respectively.

\begin{figure}[t]
    \centering
    \begin{tabular}{c}
         \includegraphics[width=\textwidth,clip]{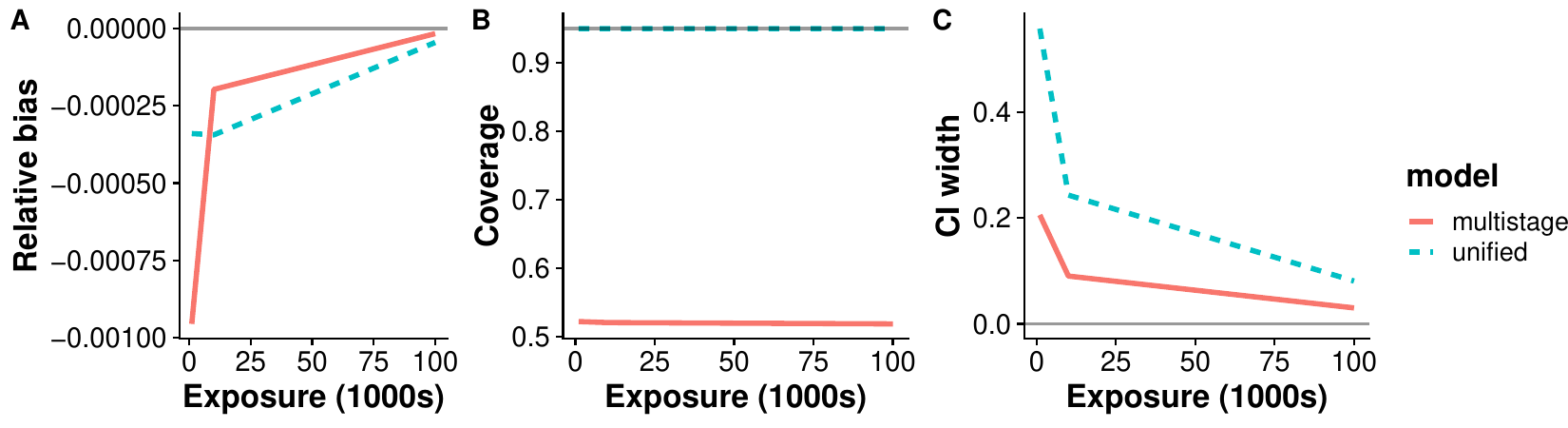}\\
         (a) Scenario 1: eight causes with extra-Poisson variability \\
         \includegraphics[width=\textwidth]{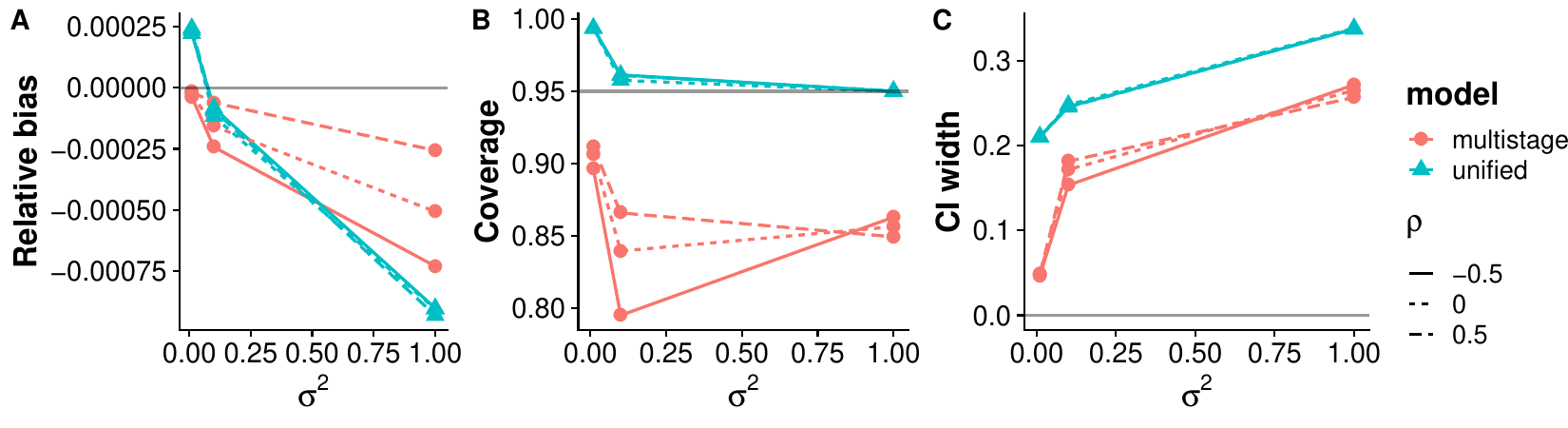} \\
         (b) Scenario 2: two correlated causes
    \end{tabular}
    \caption{Relative bias, coverage and width of 95\% intervals for log mortality rates estimates from multistage and unified models. For (a), data were generated with IID Normal random effects for each observation and three possible exposure values. For (b), data were generated with bivariate IID Normal random effects for each region-age-year strata, defining the diagonal elements of the covariance matrix as $\sigma^2$ and the off diagonal elements as $\rho\sigma^2$. Estimates are averaged over all observations and simulations per scenario.}
    \label{fig:simres}
\end{figure}

We compare the relative bias, coverage and width of posterior 95\% intervals for log mortality rates as functions of exposure time in Figure \ref{fig:simres}. Neither approach is biased but the unified model has better coverage due to appropriately wider uncertainty intervals. The problem is likelihood misspecification, because the multistage model parameterizes the log of the sums of mortality rates as normally distributed, whereas in the generated data, the sum of the log mortality rates are normally distributed, which the unified model specifies correctly. This substantial undercoverage in the multistage model is derived analytically in the supplementary material. While we are imposing a normal distribution on the log cause-specific mortality rates in this example, which may not reflect the real world, we believe this parametric form is a reasonable assumption since it corresponds to the canonical assumptions of exponential failure times in competing risks survival analysis.

\subsection{Scenario 2: correlated causes}
\label{sim2}

To show the benefits of our flexible framework for a contextually relevant situation in which a multistage approach fails, we simulate data for two causes with correlated CSMRs, representing causes with similar underlying drivers not captured in the data. 

We use the same data generating mechanism as Scenario 1 with two changes. First, we include region and age as covariates with omitted reference groups and all associated coefficients equal to $0.5$, in order to more closely resemble a situation where mortality depends on region and age. These again yield death counts and an overall mean mortality rate similar to the MCHSS data. Second, we specify $\bm\epsilon_{h} = (\epsilon_{h1}, \epsilon_{h2}) \underset{\text{iid}}{\sim} \mbox{Normal}_2(\mathbf{0},\bm\Sigma)$ for the two causes. We set the diagonal terms of $\bm\Sigma$ to be equal to $\sigma^2$, which controls overdispersion. The off-diagonal term is $\sigma^2\rho$, where $\rho$ controls the correlation between causes. We set the exposure time to the median in the MCHSS. 

We fit the same multistage model as the previous simulation, except both stages also include fixed effects for region and age, and compare it to a correctly specified unified model. We perform 100 simulations for each of nine scenarios: $\sigma^2 \in \{0.01, 0.1, 1 \}$ crossed with $\rho \in \{-0.5, 0, 0.5 \}$. The values of $\sigma^2$ span a range that are much smaller, similar to, and much larger than that estimated from the MCHSS data, and the values of $\rho$ span a range that are consistent with residual correlations estimated from the MCHSS data in an exploratory analysis in the supplementary material.

Relative bias, coverage and width of 95\% intervals for estimated log mortality rates as functions of $\sigma^2$ for each value of $\rho$ are shown in Figure \ref{fig:simres}. The bias in both models is negligible, although it increases slightly with higher overdispersion due to small-sample bias. We also see little direct impact of $\rho$. The coverage of the multistage model ranges between 80\% and 90\%, while the unified model has coverage near the nominal level with appropriately wider uncertainty intervals, although it displays overcoverage with small levels of overdispersion. This is not surprising because the elements of $\Sigma$ are all nearly 0 in this case, so our model is estimating parameters for which there is very little information. This leads to overcoverage from additional uncertainty. The multistage model's undercoverage is again due to likelihood misspecification. In summary, the flexibility of our framework facilitates modeling aspects of the data, such as overdispersion and correlation, that a multistage model cannot, which leads to more accurate inference.

\section{Estimating mortality from SRS data}
\label{china}

This section uses our framework to develop a model to estimate child mortality from the MCHSS. 

\subsection{Model description}
\label{model_description}

We index region by $r \in \{1, \dots, R=6\}$, age group by $a \in \{1, \dots, A=6\}$, year by $t \in \{1, \dots, T=20\}$, and cause by $c \in \{1, \dots, C=8\}$. Let $N_{r,a,t}$ be person-years and $y_{r,a,t,c}$ be death counts due to cause $c$ in region $r$, age group $a$, and year $t$. To estimate cause-specific mortality rates by region and age over time, we specify the model as

\begin{align}
    y_{r,a,t,c} | N_{r,a,t}, \lambda_{r,a,t,c} \sim & \text{ } \mbox{Poisson}(N_{r,a,t} \lambda_{r,a,t,c}) \nonumber \\
    \log\left(\lambda_{r,a,t,c}\right) = &\text{ } \alpha + \beta^R_r + \beta^A_a + \beta^C_c + \label{eq:meanmodel} \\
    &\text{ } \beta^{AC}_{a,c} + \beta^{RC}_{r,c} + \beta^{AR}_{a,r} + \nonumber \\
    &\text{ } \gamma_{r,a^{\star}[a],c^{\star}[c]}(t) + \epsilon_{r,a,t,c} \nonumber
\end{align}

In order to properly use a Poisson likelihood, we rounded deaths up to the nearest integer because the death counts in the MCHSS were fractional due to the underreporting adjustment described in Section \ref{data}. 

In (\ref{eq:meanmodel}), $\alpha$ is the overall intercept, and $\beta^R_r$, $\beta^A_a$, and $\beta^C_c$ are fixed effects for region, age, and cause, respectively. These are specified with omitted reference groups, i.e. $\beta^R_1 = \beta^A_1 = \beta^C_1 = 0$. We chose to use fixed effects due to the small numbers of regions, age groups, and causes along with the strong differences in mortality among the different strata. Notably, due to regions being defined not simply by geography but by urban/rural status as well, we did not include any spatially dependent effects because there were only three geographic regions.

Notably, we did not model correlations between causes. We performed a simulation study, presented in the supplementary material and further discussed in Section \ref{discussion}, in which we simulated two data sets with correlated CSMRs---one that was the same size as the MCHSS and one with 100 regions rather than six. The correlation parameters were well-estimated for the data with 100 regions but not for the MCHSS-sized data. Consequently, we did not model correlations in our final model.

To capture first-order interactions among cause, age, and region, we include $\beta^{AC}_{a,c}$, $\beta^{RC}_{r,c}$, and $\beta^{AR}_{a,r}$ as fixed effects with omitted reference groups. To determine how to model these non-temporal interactions, initial data exploration using non-Bayesian GLMs showed that a generalized linear model with all three of these interactions had a substantially lower AIC than any model with only one of them included, and had only slightly higher AIC than a model that additionally included a three-way interaction. Because the model with a three-way interaction had approximately twice as many parameters, we chose to only include two-way interactions for parsimony. 

We place proper but flat priors on all the above fixed effects except $\alpha$ for which we use an improper flat prior due to our choice to model temporal trends with a second-order random walk (discussed later). Prior choice is important in more data sparse situations, but here estimating with stronger priors makes little difference, as we establish with sensitivity analyses described in Section \ref{discussion}. 

The parameter $\gamma_{r,a^{\star}[a],c^{\star}[c]}(t)$ is a random effect on time with a second-order random walk distribution, which we denote as $\gamma_{r,a^{\star},c^{\star}}(t) \sim \mbox{RW2}(\sigma^2_{\gamma})$. We include different random walks for various age-region-cause combinations and we share random walks among certain ages and causes, hence indexing by $a^{\star}$ and $c^{\star}$. We define $a^{\star}[a] = 1$ for observations in the 0--6 day age group ($a = 1$) and $a^{\star}[a] = 2$ otherwise ($a = 2, \dots, 6$). We define $c^{\star}[c] = 1$ for \textit{diarrhea} and \textit{other communicable diseases} ($c = 1, 2$), $c^{\star}[c] = 2$ for \textit{congenital anomalies} and \textit{other non-communicable diseases} ($c = 3, 4$), and $c^{\star}[c] = 3, \dots, 6$ for the remaining causes ($c = 5, \dots, 8$, respectively). This results in 6 (region) $\times$ 2 (age) $\times$ 6 (cause) $= 72$ random walks. By using different random walks to allow age-region-cause strata to have distinct trends, we accomplish a similar goal as the model in \cite{he2017national} without using ad-hoc weighted rolling averages. All random walks share a variance parameter, $\sigma^2_{\gamma}$ for parsimony and to reduce the number of estimated parameters to aid computation. For identifiability, we use sum-to-zero constraints on each random walk along with an improper prior on the intercept $\alpha$, which is required for correct specification of a second-order random walk \citep{rue2005gaussian}. We use a penalized complexity prior \citep{simpson2017penalising} on the variance parameters such that there is a $1\%$ probability that $\sigma_{\gamma} > 1$. 

We use a second-order random walk to encourage the estimated mortality rates to vary smoothly in time. A second-order random walk model penalizes deviations from linearity and is more smooth than a first-order random walk, which models the first differences in time as being normally distributed and allows for sharp year-to-year variation. The additional smoothness of the second-order random walk is in line with what we expect in the MCHSS data.

We share random walks among certain ages and causes to aid computation. The categories for sharing random walks were chosen via a data-driven exercise that accounted for the scientific context. We fit a suite of Poisson generalized linear models that contained interactions between time and all one-way, two-way, and three-way combinations of region, age, and cause, and then analyzed the residual plots for common patterns that were consistent with the context of child mortality in China. The 0--6 day age group had consistent patterns in the residuals that were different than the other ages, which is reasonable due to the biological uniqueness of this age group such as higher mortality and its dependence on birth-related interventions of health facilities. The causes that share random walks also had similar residual patterns which are reasonable because \textit{diarrhea} and \textit{other communicable diseases} are communicable, while \textit{congenital anomalies} and \textit{other non-communicable diseases} are non-communicable. This is fully detailed in the supplementary material. 

We tested the feasibility of the shared variance parameter by separately fitting random walk models for the data in each of the 72 age-region-cause combinations and comparing the estimated standard deviation parameters, which are presented in the supplementary material. The estimates ranged from 0.005 to 0.1, with the majority below 0.025. Sharing the variance parameter will shrink the rates of change in some of the time trends toward the average, but not drastically.

We specify $\epsilon_{r,a,t,c} \underset{\text{iid}}{\sim} \text{Normal}(0,\sigma_{\epsilon}^2)$ with a penalized complexity prior on $\sigma^2_{\epsilon}$ such that there is a $1\%$ probability that $\sigma_{\epsilon} > 5$. We include this term to account for overdispersion as discussed in Section \ref{sub_overdispersion}. We treat $\epsilon_{r,a,t,c}$ as an error term rather than true signal and do not include it in the posterior distribution of our final estimates. While this parameter likely captures some true signal, we believe the relative strength of the signal is low because the data is a sample with quality issues, which we discuss in Section \ref{discussion}. With more covariates and higher data quality, the magnitude of the noise component would decrease. Furthermore, by omitting $\epsilon_{r,a,t,c}$, our final estimates reflect the underlying smooth time trends. We may wish to include this parameter in contexts with more data, higher data quality, or where the goal is to estimate true numbers of deaths in a population rather than underlying mortality rates.

To demonstrate model adequacy, we plotted $(y_{r,a,t,c} - N_{r,a,t} \widehat\lambda_{r,a,t,c}) / (N_{r,a,t} \widehat\lambda_{r,a,t,c})^{1/2}$, the standardized residuals, by all two- and three-way combinations of age, region, cause, and time. These plots were examined for patterns that may suggest inadequate model fit to the data. As an example of evaluating predictive performance, we held out the final year of data, fit the model, predicted the log mortality rates, and then plotted these against the held out data. This particular hold-out experiment was performed because short-term predictions are relevant for policy decisions.  Finally for model comparison, we fit a model with no interactions between fixed effects to assess the necessity of these interactions, and a model using first order autoregressive (AR1) processes rather than second-order random walks to assess how we modeled temporal trends. We compared these to our final model via DIC, WAIC, and negative sum of the log CPO.

We fit the model with the \texttt{INLA} package in \texttt{R}. Code for data processing, model fitting, and plotting results is available at \href{http://www.github.com/aeschuma/SRS-child-mortality}{http://www.github.com/aeschuma/SRS-child-mortality}.

\begin{figure}[t]
    \centering
    \begin{tabular}{c}
         \includegraphics[page=24, width=0.95\textwidth, trim=0in 4.5in 0in 0in, clip]{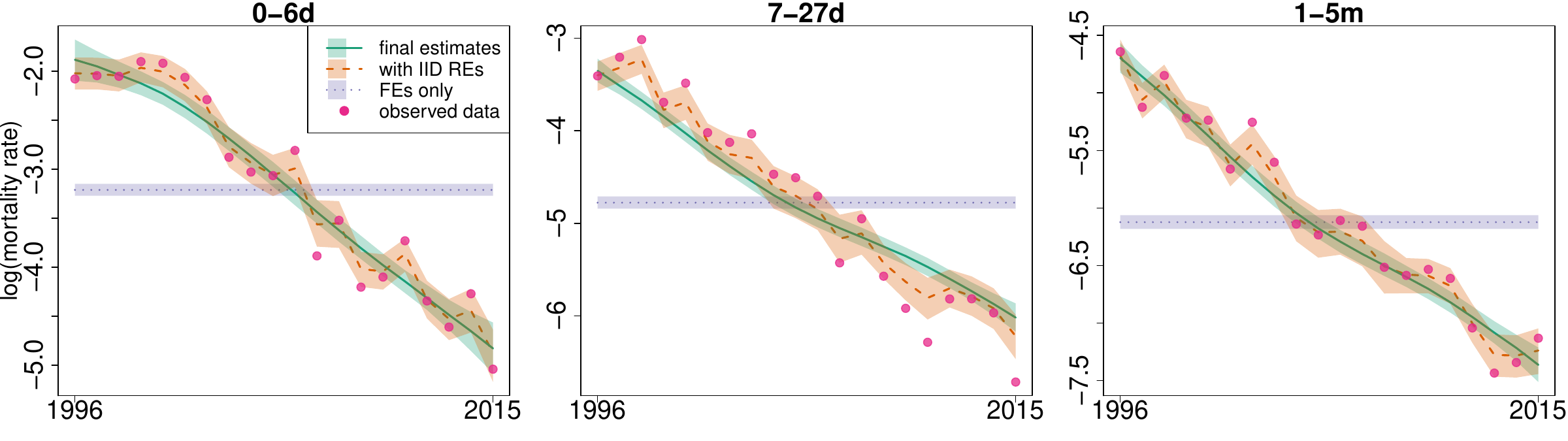} \\
         (a) prematurity in the west rural region \\  \includegraphics[page=28,width=0.95\textwidth]{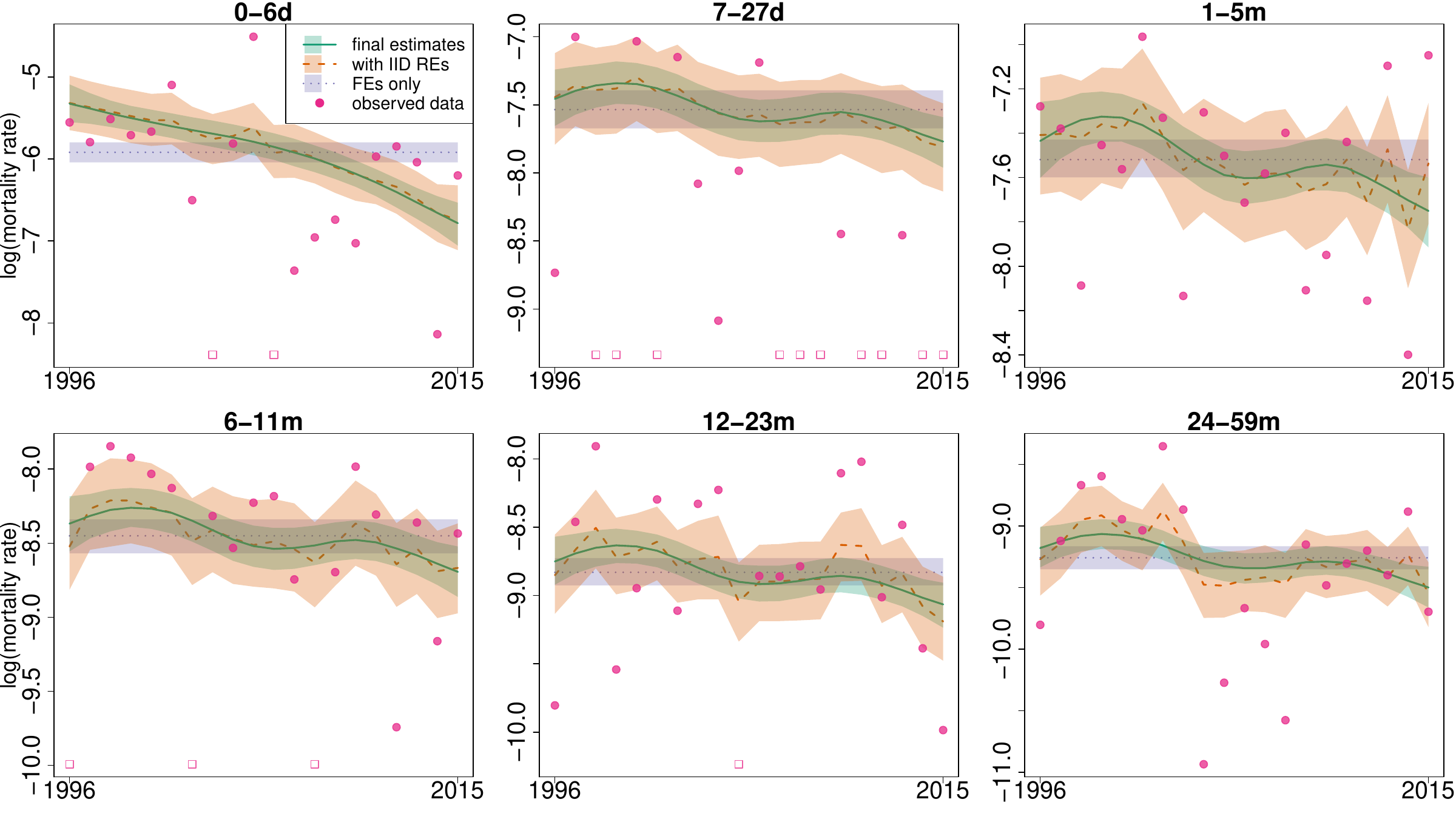} \\
         (b) other non-communicable diseases in the east urban region  
    \end{tabular}
    \caption{Selected results from the MCHSS data showing empirical data, estimated posterior medians, and posterior 80\% intervals for log mortality rates. Combinations with no deaths are represented by an open square.}
    \label{fig:logmxres}
\end{figure}

\subsection{Results}
\label{china_results}

Figure \ref{fig:logmxres} shows estimated posterior medians and posterior 80\% intervals for the log mortality rates over time in each age group for selected regions and causes. In order to show how the fixed effects component of the model contributes to these estimates, we include posterior medians and 80\% intervals for the sum of the fixed effects only; to see the contribution of the random effect error terms, $\epsilon_{r,a,t,c}$, we also show the posterior medians and 80\% intervals for the estimated log mortality rates with these added. As discussed in Section \ref{model_description}, there is ambiguity in the source of these error terms, but their inclusion allows for a greater sense of the sampling variation in the data to be realized. 

We first present estimates for prematurity in the west rural region, which represents the highest number of deaths in the data. Our model fits well, although the estimates are consistently higher than the observed data in the 7--27 day age group in later years, which is due to borrowing strength across other strata. This may indicate data errors, such as underreporting missed by the adjustment, or it may be due to shrinkage induced by the random walks. In comparison to preliminary models with random walks for each age-region combination only, the time trends for west rural prematurity were carried over to all causes and did not fit the data well. This is testament to the importance of including random walks by age-region-cause strata. Looking at a different cause and region, other non-communicable diseases in the east urban region, we see largely flatter time trends with wider uncertainty reflecting the smaller amount of data. In the random walk fitting exercise described in Section \ref{model_description}, this strata had a much lower estimated random walk variance parameter than the previous two strata presented, but all have acceptable fits here. Plots of all estimates from our final model are available in the supplementary material. Additionally in the supplementary material, we present estimated CSMFs over time for each age group and region to show differences in the disribution of causes by age and region over time. For example, the percentage of deaths due to congenital anomalies in the 1-5 and 6-11 month age groups is fairly constant in the east urban region but increasing in the mid rural region. 

To show model adequacy, we present plots of standardized residuals stratified by all two- and three-way combinations of age, region, cause, and time, and plots of predicted vs. observed from a model with the final year held out. These show no gross misspecification and predictions with no systematic biases. Furthermore, in the supplementary material we present comparisons of the DIC, WAIC, and negative sum of the log CPO among our final model, a model with no interactions between fixed effects, and a model using AR1 processes rather than second-order random walks. Not surprisingly, leaving out interactions caused a substantially poorer fit; the AR1 model was close but still inferior.

\begin{figure}[t]
    \centering
    \includegraphics[width=0.95\textwidth]{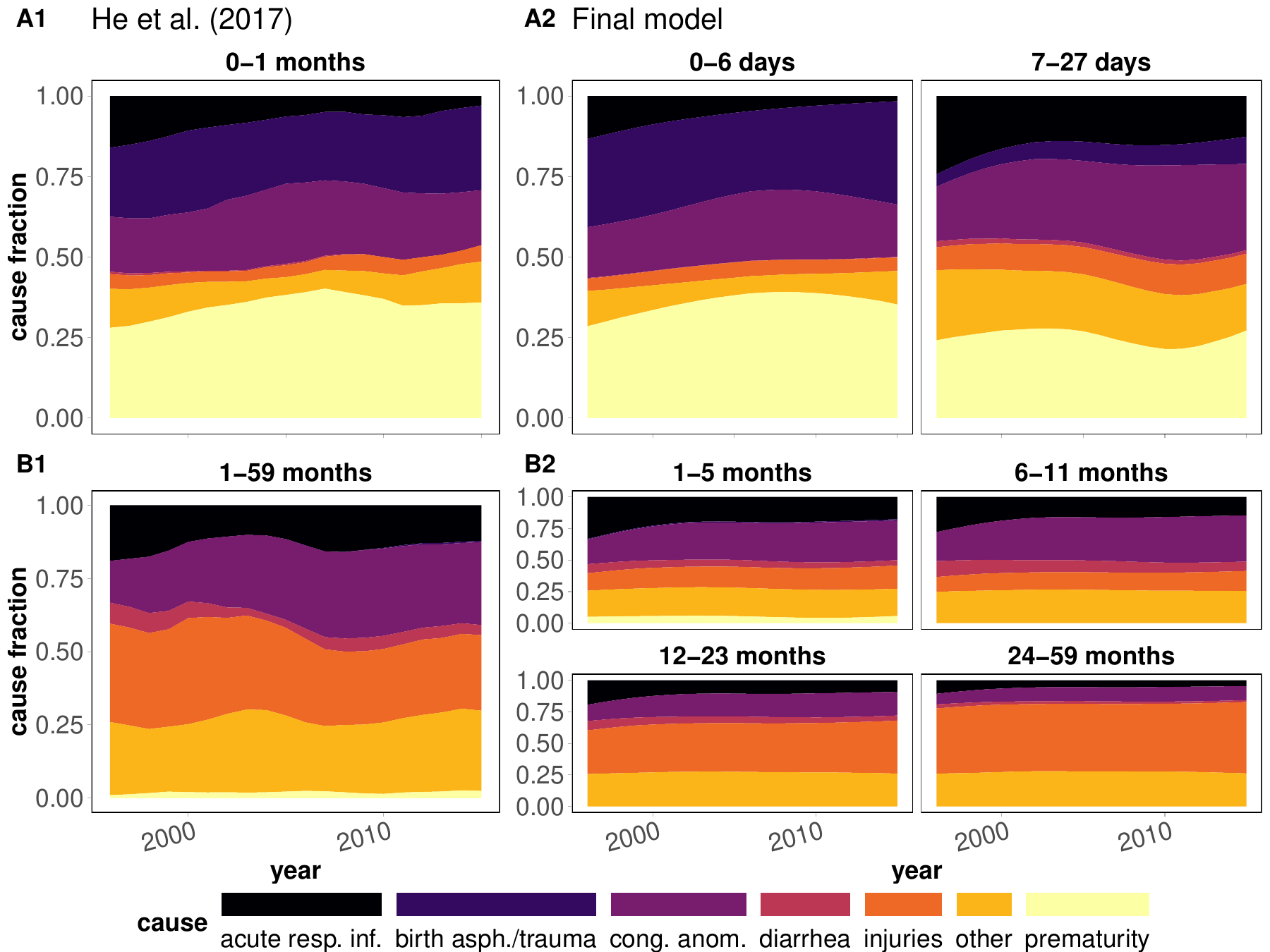}
    \caption{Comparisons of estimated CSMFs between our model and the model in \cite{he2017national} in the east rural region. Panels A1 and B1 show cause fractions from \cite{he2017national} for the 0--1 month and 1--59 month age groups, respectively, while panels A2 and B2 show the estimates from our model for the finer age groups as labeled that are within the broad age groups in A1 and B1.}
    \label{fig:hecompare}
\end{figure}

Since the method of \cite{he2017national} also estimates cause-specific child mortality from the MCHSS data, we compare their results to our estimated CSMFs in the east rural region as an example in Figure \ref{fig:hecompare}. One primary benefit of our estimates is the more detailed age granularity. The approach in \cite{he2017national} produced estimates for the 0--1 month and 1--59 month age groups. Any further disaggregation would presumably lead to unstable estimation as their model performs temporal smoothing from moving averages for each region-age-cause separately. In comparison, our method borrows strength in two key ways in order to facilitate modeling more granular age groups. First, we include fixed effects for interactions among age groups, causes of death, and regions. Second, we share random walks in time among different strata. Our estimates show substantial heterogeneity across the age groups that is masked by the large age bins in \cite{he2017national}. Figure \ref{fig:hecompare} also shows smoother temporal trends in our estimates, demonstrating less year-to-year variability which is what we expect for the population-level parameters in China that we are estimating from our sample.

This comparison illustrates important conclusions that are missed by \cite{he2017national}. In the 0--1 month age group, \cite{he2017national} estimate a large decline in the cause fraction for acute respiratory infections, resulting in a near-negligible percentage in 2015. However, our model shows that this cause is still meaningful in the 7--27 day age group. For intervention funding allocated in this region, our results indicate continued investment in acute respiratory infection prevention among children aged 7-27 whereas the results from \cite{he2017national} do not. Another key takeaway is the comparatively larger portion of deaths due to injuries in the 24--59 month age group, along with smaller portions due to acute respiratory infections and congenital anomalies. This pattern could influence policies and resources to be enacted in a more targeted manner.

\section{Discussion}
\label{discussion}

We have introduced a unified, flexible framework for estimating age- and cause-specific child mortality over time using tabulated death counts and exposure time from SRS data. This framework is based on an individual-level competing risks likelihood along with Bayesian smoothing priors. We have shown that it performs better than multistage modeling on simulated data with overdispersion and correlation, and we used the framework to develop a model for the MCHSS data from China. 

Our framework improves upon current methodologies by simultaneously estimating all-cause and cause-specific mortality in a unified framework rather than using improper multistage models. We compared our model for the MCHSS data in China to the model in \cite{he2017national}. Our model improved estimation in four key ways: (1) \cite{he2017national} improperly scales to national estimates as part of their multistage model, which is effectively unnecessary in our model because all-cause and cause-specific estimates are estimated simultaneously and consistently, which properly quantifies uncertainty; (2) we directly derive a likelihood for tabulated death and exposure time data commonly reported from SRS which allows us to fully specify a flexible estimation model in a statistically well-grounded fashion; (3) we model temporal trends with second-order random walks rather than using ad hoc moving averages;  and (4) we employ proper smoothing and variance estimation that allows stable estimates at finer subgroup granularity. Our estimates are smoother and tease out important heterogeneity in finer age groups that is missed in the results from \cite{he2017national}.

While we developed our framework to estimate child mortality from SRS data, it is applicable to other scenarios. We can use cause-specific child mortality data tabulated from household surveys (as long as sampling probabilities are correctly accounted for), although modeling the individual data in this case may be preferred as discussed below. Additionally, high quality VR can be modeled with our framework, for which we recommend less smoothing due to near complete population coverage. Furthermore, by incorporating multiple types of data from multiple countries, our framework can be adapted for large scale estimation akin to \cite{liu2016global} and \cite{vos2020global}. This extension would be computationally expensive, however, necessitating developments to be explored in future research. Beyond child mortality, we can use our framework to develop models for estimating any rates with competing risks from data sources that provide tabulated counts and time-at-risk, or from which these values can be calculated. Some examples are cause-specific mortality for ages beyond 5 years, incidence of non-fatal diseases, and rates of traffic accidents by severity.

The high degree of flexibility in our framework allows complex trends to be estimated, but data availability should drive model choice. Simulation experiments with data generated similar to the true data can reveal what forms are estimable. As one example, we attempted an alternate specification for the MCHSS data that accounted for correlations between causes. We simulated data with the same size as the MCHSS that had correlated CSMRs and fit a correctly specified model. Recovery of the correlation parameters was poor and posterior distributions were wide. We then simulated data with 100 regions rather than six and fit a similar model. The correlation parameters were recovered well with narrow posterior distributions. Thus, we chose not to model correlation. A full description of this simulation exercise and its results are provided in the supplementary material. 

Additionally, choosing prior distributions requires careful thought. We used diffuse priors for the fixed effects and penalized complexity priors for the variance parameters of the random walks and IID Normal random effects. The latter parameters are more sensitive to prior choice. We recommend penalized complexity priors due to the reasons outlined in \cite{simpson2017penalising}, in particular allowing for specifying context-relevant prior distributions. As sensitivity analyses for our model fit to MCHSS data, we fit one model with Gamma(5, 0.00005) priors on the precisions and one model with stronger Normal(0,5) priors on the fixed effects. No appreciable differences were found; the largest absolute difference in posterior median log mortality rates was 0.05.

We propose that the data context suggests the most important aspects of child mortality to model for the setting at hand. From this viewpoint, one could fit multiple candidate models and perform model assessment to choose the most suitable model for the scientific goal. The independent random effects are useful for this endeavor, for example plotting them against time to detect patterns. Cross validation is also useful and could include multiple levels depending on the context (e.g. leaving one observation out, leaving one region out, leaving one time period out). Different levels of cross validation explore aspects of the data for which different models may perform better \citep{roberts2017cross}.

Finally, the MCHSS data provide an example of common problems to be addressed in SRS data. Underreporting adjustment and exposure time estimation are common data preprocessing steps. Beyond this, errors in cause attribution introduce substantial variability \citep{desai2014performance,murray2014using}. These issues naturally lead to using random effects in order to induce overdispersion which may accommodate this extra variability, though they cannot of course remove bias. If available, the unadjusted, unaggregated data would be used and the aggregation and completeness adjustments would be included as components of the Bayesian model. Using individual-level data allows proper calculation of person-time, necessitating proper handling of the censoring for individual-level survival data. Individual-level data also require explicit incorporation of sampling probabilities either with design-based or model-based estimation \citep{pfeffermann2013new}. With individual-level VA data, probabilistic cause assignment could also be included, for example using the method in \cite{mccormick2016probabilistic}. Extending our model in this manner would be necessary when using smaller scale surveillance data on individuals, e.g. data from COMSA Mozambique \citep{nkengasong2020improving}, HDSS sites in the INDEPTH network \citep{sankoh2012indepth}, or HDSS sites in the ALPHA network \citep{maher2010translating}. Future work could also expand this framework to include survey data, notably VA data, in a single model.

\begin{supplement}
\stitle{Supplementary material for ``A flexible Bayesian framework to estimate age- and cause-specific child mortality over time from sample registration data''} \slink[url]{To be published}
\sdescription{The supplementary material provides an overview of SRS and associated data, an explanation of the method to calculate exposure time in the MCHSS, summary statistics for the MCHSS data, the full derivation of the proposed likelihood in Section \ref{theory}, full descriptions of the simulations in Section \ref{sims} and additional analytical derivations of result, other simulations mentioned in Sections \ref{sims} and \ref{discussion}, descriptions and results of the model development exercises and analyses using the mortality estimates discussed in Section \ref{china}, graphs of all log mortality rate estimates from the MCHSS data from the final model, a graph of cause-specific mortality fractions over time by age and region, and plots demonstrating the fit and adequacy of the final model.}
\end{supplement}

\bibliography{mybib}

\begin{thebibliography}{68}

\bibitem[\protect\citeauthoryear{Abdullah et~al.}{2007}]{abdullah2007patterns}
\begin{barticle}[author]
\bauthor{\bsnm{Abdullah},~\bfnm{Salim}\binits{S.}},
  \bauthor{\bsnm{Adazu},~\bfnm{Kubaje}\binits{K.}},
  \bauthor{\bsnm{Masanja},~\bfnm{Honorati}\binits{H.}},
  \bauthor{\bsnm{Diallo},~\bfnm{Diadier}\binits{D.}},
  \bauthor{\bsnm{Hodgson},~\bfnm{Abraham}\binits{A.}},
  \bauthor{\bsnm{Ilboudo-Sanogo},~\bfnm{Edith}\binits{E.}},
  \bauthor{\bsnm{Nhacolo},~\bfnm{Ariel}\binits{A.}},
  \bauthor{\bsnm{Owusu-Agyei},~\bfnm{Seth}\binits{S.}},
  \bauthor{\bsnm{Thompson},~\bfnm{Ricardo}\binits{R.}},
  \bauthor{\bsnm{Smith},~\bfnm{Thomas}\binits{T.}} \betal{et~al.}
(\byear{2007}).
\btitle{Patterns of age-specific mortality in children in endemic areas of
  sub-Saharan Africa.}
\bjournal{The American Journal of Tropical Medicine and Hygiene}
\bvolume{77}
\bpages{99--105}.
\end{barticle}
\endbibitem

\bibitem[\protect\citeauthoryear{AbouZahr et~al.}{2015a}]{abouzahr2015towards}
\begin{barticle}[author]
\bauthor{\bsnm{AbouZahr},~\bfnm{Carla}\binits{C.}},
  \bauthor{\bsnm{De~Savigny},~\bfnm{Don}\binits{D.}},
  \bauthor{\bsnm{Mikkelsen},~\bfnm{Lene}\binits{L.}},
  \bauthor{\bsnm{Setel},~\bfnm{Philip~W}\binits{P.~W.}},
  \bauthor{\bsnm{Lozano},~\bfnm{Rafael}\binits{R.}} \AND
  \bauthor{\bsnm{Lopez},~\bfnm{Alan~D}\binits{A.~D.}}
(\byear{2015}a).
\btitle{Towards universal civil registration and vital statistics systems: the
  time is now}.
\bjournal{The Lancet}
\bvolume{386}
\bpages{1407--1418}.
\end{barticle}
\endbibitem

\bibitem[\protect\citeauthoryear{AbouZahr et~al.}{2015b}]{abouzahr2015civil}
\begin{barticle}[author]
\bauthor{\bsnm{AbouZahr},~\bfnm{Carla}\binits{C.}},
  \bauthor{\bsnm{De~Savigny},~\bfnm{Don}\binits{D.}},
  \bauthor{\bsnm{Mikkelsen},~\bfnm{Lene}\binits{L.}},
  \bauthor{\bsnm{Setel},~\bfnm{Philip~W}\binits{P.~W.}},
  \bauthor{\bsnm{Lozano},~\bfnm{Rafael}\binits{R.}},
  \bauthor{\bsnm{Nichols},~\bfnm{Erin}\binits{E.}},
  \bauthor{\bsnm{Notzon},~\bfnm{Francis}\binits{F.}} \AND
  \bauthor{\bsnm{Lopez},~\bfnm{Alan~D}\binits{A.~D.}}
(\byear{2015}b).
\btitle{Civil registration and vital statistics: progress in the data
  revolution for counting and accountability}.
\bjournal{The Lancet}
\bvolume{386}
\bpages{1373--1385}.
\end{barticle}
\endbibitem

\bibitem[\protect\citeauthoryear{{United Nations Inter-agency Group for Child
  Mortality Estimation}}{2020}]{igme2020levels}
\begin{btechreport}[author]
\bauthor{\bsnm{{United Nations Inter-agency Group for Child Mortality
  Estimation}}}
(\byear{2020}).
\btitle{Levels \& Trends in Child Mortality 2020}
\btype{Technical Report},
\bpublisher{United Nations Children's Fund}.
\end{btechreport}
\endbibitem

\bibitem[\protect\citeauthoryear{Alkema and New}{2014}]{alkema2014global}
\begin{barticle}[author]
\bauthor{\bsnm{Alkema},~\bfnm{Leontine}\binits{L.}} \AND
  \bauthor{\bsnm{New},~\bfnm{Jin~Rou}\binits{J.~R.}}
(\byear{2014}).
\btitle{Global estimation of child mortality using a {B}ayesian {B}-spline
  bias-reduction model}.
\bjournal{The Annals of Applied Statistics}
\bpages{2122--2149}.
\end{barticle}
\endbibitem

\bibitem[\protect\citeauthoryear{Aponte et~al.}{2009}]{aponte2009efficacy}
\begin{barticle}[author]
\bauthor{\bsnm{Aponte},~\bfnm{John~J}\binits{J.~J.}},
  \bauthor{\bsnm{Schellenberg},~\bfnm{David}\binits{D.}},
  \bauthor{\bsnm{Egan},~\bfnm{Andrea}\binits{A.}},
  \bauthor{\bsnm{Breckenridge},~\bfnm{Alasdair}\binits{A.}},
  \bauthor{\bsnm{Carneiro},~\bfnm{Ilona}\binits{I.}},
  \bauthor{\bsnm{Critchley},~\bfnm{Julia}\binits{J.}},
  \bauthor{\bsnm{Danquah},~\bfnm{Ina}\binits{I.}},
  \bauthor{\bsnm{Dodoo},~\bfnm{Alexander}\binits{A.}},
  \bauthor{\bsnm{Kobbe},~\bfnm{Robin}\binits{R.}},
  \bauthor{\bsnm{Lell},~\bfnm{Bertrand}\binits{B.}} \betal{et~al.}
(\byear{2009}).
\btitle{Efficacy and safety of intermittent preventive treatment with
  sulfadoxine-pyrimethamine for malaria in {A}frican infants: a pooled analysis
  of six randomised, placebo-controlled trials}.
\bjournal{The Lancet}
\bvolume{374}
\bpages{1533--1542}.
\end{barticle}
\endbibitem

\bibitem[\protect\citeauthoryear{Bchir et~al.}{2006}]{bchir2006better}
\begin{barticle}[author]
\bauthor{\bsnm{Bchir},~\bfnm{Abdallah}\binits{A.}},
  \bauthor{\bsnm{Bhutta},~\bfnm{Zulfiqar}\binits{Z.}},
  \bauthor{\bsnm{Binka},~\bfnm{Fred}\binits{F.}},
  \bauthor{\bsnm{Black},~\bfnm{Robert}\binits{R.}},
  \bauthor{\bsnm{Bradshaw},~\bfnm{Debbie}\binits{D.}},
  \bauthor{\bsnm{Garnett},~\bfnm{Geoff}\binits{G.}},
  \bauthor{\bsnm{Hayashi},~\bfnm{Kenji}\binits{K.}},
  \bauthor{\bsnm{Jha},~\bfnm{Prabhat}\binits{P.}},
  \bauthor{\bsnm{Peto},~\bfnm{Richard}\binits{R.}},
  \bauthor{\bsnm{Sawyer},~\bfnm{Cheryl}\binits{C.}} \betal{et~al.}
(\byear{2006}).
\btitle{Better health statistics are possible}.
\bjournal{The Lancet}
\bvolume{367}
\bpages{190--193}.
\end{barticle}
\endbibitem

\bibitem[\protect\citeauthoryear{Bennett and
  Wakefield}{2001}]{bennett2001errors}
\begin{barticle}[author]
\bauthor{\bsnm{Bennett},~\bfnm{James}\binits{J.}} \AND
  \bauthor{\bsnm{Wakefield},~\bfnm{Jon}\binits{J.}}
(\byear{2001}).
\btitle{Errors-in-variables in joint population pharmacokinetic/pharmacodynamic
  modeling}.
\bjournal{Biometrics}
\bvolume{57}
\bpages{803--812}.
\end{barticle}
\endbibitem

\bibitem[\protect\citeauthoryear{Boerma}{2013}]{boerma2013public}
\begin{barticle}[author]
\bauthor{\bsnm{Boerma},~\bfnm{J~Ties}\binits{J.~T.}}
(\byear{2013}).
\btitle{Public health information needs in districts}.
\bjournal{BMC {H}ealth {S}ervices {R}esearch}
\bvolume{13}
\bpages{S12}.
\end{barticle}
\endbibitem

\bibitem[\protect\citeauthoryear{Boerma and
  Stansfield}{2007}]{boerma2007health}
\begin{barticle}[author]
\bauthor{\bsnm{Boerma},~\bfnm{J~Ties}\binits{J.~T.}} \AND
  \bauthor{\bsnm{Stansfield},~\bfnm{Sally~K}\binits{S.~K.}}
(\byear{2007}).
\btitle{Health statistics now: are we making the right investments?}
\bjournal{The Lancet}
\bvolume{369}
\bpages{779--786}.
\end{barticle}
\endbibitem

\bibitem[\protect\citeauthoryear{Breslow and
  Clayton}{1993}]{breslow1993approximate}
\begin{barticle}[author]
\bauthor{\bsnm{Breslow},~\bfnm{Norman~E}\binits{N.~E.}} \AND
  \bauthor{\bsnm{Clayton},~\bfnm{David~G}\binits{D.~G.}}
(\byear{1993}).
\btitle{Approximate inference in generalized linear mixed models}.
\bjournal{Journal of the American Statistical Association}
\bvolume{88}
\bpages{9--25}.
\end{barticle}
\endbibitem

\bibitem[\protect\citeauthoryear{Chi and Reinsel}{1989}]{chi1989models}
\begin{barticle}[author]
\bauthor{\bsnm{Chi},~\bfnm{Eric~M}\binits{E.~M.}} \AND
  \bauthor{\bsnm{Reinsel},~\bfnm{Gregory~C}\binits{G.~C.}}
(\byear{1989}).
\btitle{Models for longitudinal data with random effects and AR (1) errors}.
\bjournal{Journal of the American Statistical Association}
\bvolume{84}
\bpages{452--459}.
\end{barticle}
\endbibitem

\bibitem[\protect\citeauthoryear{Clark, Setel and Li}{2019}]{clark2019verbal}
\begin{barticle}[author]
\bauthor{\bsnm{Clark},~\bfnm{Samuel~J}\binits{S.~J.}},
  \bauthor{\bsnm{Setel},~\bfnm{Philip}\binits{P.}} \AND
  \bauthor{\bsnm{Li},~\bfnm{Zehang}\binits{Z.}}
(\byear{2019}).
\btitle{Verbal Autopsy in Civil Registration and Vital Statistics: The
  Symptom-Cause Information Archive}.
\bjournal{arXiv preprint arXiv:1910.00405}.
\end{barticle}
\endbibitem

\bibitem[\protect\citeauthoryear{Datta et~al.}{2016}]{datta2016nonseparable}
\begin{barticle}[author]
\bauthor{\bsnm{Datta},~\bfnm{Abhirup}\binits{A.}},
  \bauthor{\bsnm{Banerjee},~\bfnm{Sudipto}\binits{S.}},
  \bauthor{\bsnm{Finley},~\bfnm{Andrew~O}\binits{A.~O.}},
  \bauthor{\bsnm{Hamm},~\bfnm{Nicholas~AS}\binits{N.~A.}} \AND
  \bauthor{\bsnm{Schaap},~\bfnm{Martijn}\binits{M.}}
(\byear{2016}).
\btitle{Nonseparable dynamic nearest neighbor Gaussian process models for large
  spatio-temporal data with an application to particulate matter analysis}.
\bjournal{The annals of applied statistics}
\bvolume{10}
\bpages{1286}.
\end{barticle}
\endbibitem

\bibitem[\protect\citeauthoryear{Datta et~al.}{2019}]{datta2019spatial}
\begin{barticle}[author]
\bauthor{\bsnm{Datta},~\bfnm{Abhirup}\binits{A.}},
  \bauthor{\bsnm{Banerjee},~\bfnm{Sudipto}\binits{S.}},
  \bauthor{\bsnm{Hodges},~\bfnm{James~S}\binits{J.~S.}},
  \bauthor{\bsnm{Gao},~\bfnm{Leiwen}\binits{L.}} \betal{et~al.}
(\byear{2019}).
\btitle{Spatial disease mapping using directed acyclic graph auto-regressive
  (DAGAR) models}.
\bjournal{Bayesian Analysis}
\bvolume{14}
\bpages{1221--1244}.
\end{barticle}
\endbibitem

\bibitem[\protect\citeauthoryear{Desai et~al.}{2014}]{desai2014performance}
\begin{barticle}[author]
\bauthor{\bsnm{Desai},~\bfnm{Nikita}\binits{N.}},
  \bauthor{\bsnm{Aleksandrowicz},~\bfnm{Lukasz}\binits{L.}},
  \bauthor{\bsnm{Miasnikof},~\bfnm{Pierre}\binits{P.}},
  \bauthor{\bsnm{Lu},~\bfnm{Ying}\binits{Y.}},
  \bauthor{\bsnm{Leitao},~\bfnm{Jordana}\binits{J.}},
  \bauthor{\bsnm{Byass},~\bfnm{Peter}\binits{P.}},
  \bauthor{\bsnm{Tollman},~\bfnm{Stephen}\binits{S.}},
  \bauthor{\bsnm{Mee},~\bfnm{Paul}\binits{P.}},
  \bauthor{\bsnm{Alam},~\bfnm{Dewan}\binits{D.}},
  \bauthor{\bsnm{Rathi},~\bfnm{Suresh~Kumar}\binits{S.~K.}} \betal{et~al.}
(\byear{2014}).
\btitle{Performance of four computer-coded verbal autopsy methods for cause of
  death assignment compared with physician coding on 24,000 deaths in low-and
  middle-income countries}.
\bjournal{BMC {M}edicine}
\bvolume{12}
\bpages{20}.
\end{barticle}
\endbibitem

\bibitem[\protect\citeauthoryear{{United Nations Population
  Division}}{2019}]{united2019world}
\begin{btechreport}[author]
\bauthor{\bsnm{{United Nations Population Division}}}
(\byear{2019}).
\btitle{{World Population Prospects}}
\btype{Technical Report},
\bpublisher{Dept of {I}nternational {E}conomic and {S}ocial {A}ffairs}.
\end{btechreport}
\endbibitem

\bibitem[\protect\citeauthoryear{Friberg et~al.}{2010}]{friberg2010sub}
\begin{barticle}[author]
\bauthor{\bsnm{Friberg},~\bfnm{Ingrid~K}\binits{I.~K.}},
  \bauthor{\bsnm{Kinney},~\bfnm{Mary~V}\binits{M.~V.}},
  \bauthor{\bsnm{Lawn},~\bfnm{Joy~E}\binits{J.~E.}},
  \bauthor{\bsnm{Kerber},~\bfnm{Kate~J}\binits{K.~J.}},
  \bauthor{\bsnm{Odubanjo},~\bfnm{M~Oladoyin}\binits{M.~O.}},
  \bauthor{\bsnm{Bergh},~\bfnm{Anne-Marie}\binits{A.-M.}},
  \bauthor{\bsnm{Walker},~\bfnm{Neff}\binits{N.}},
  \bauthor{\bsnm{Weissman},~\bfnm{Eva}\binits{E.}},
  \bauthor{\bsnm{Chopra},~\bfnm{Mickey}\binits{M.}},
  \bauthor{\bsnm{Black},~\bfnm{Robert~E}\binits{R.~E.}} \betal{et~al.}
(\byear{2010}).
\btitle{Sub-Saharan Africa's mothers, newborns, and children: how many lives
  could be saved with targeted health interventions?}
\bjournal{PLoS Medicine}
\bvolume{7}
\bpages{e1000295}.
\end{barticle}
\endbibitem

\bibitem[\protect\citeauthoryear{Gelman, Hwang and
  Vehtari}{2014}]{gelman2014understanding}
\begin{barticle}[author]
\bauthor{\bsnm{Gelman},~\bfnm{Andrew}\binits{A.}},
  \bauthor{\bsnm{Hwang},~\bfnm{Jessica}\binits{J.}} \AND
  \bauthor{\bsnm{Vehtari},~\bfnm{Aki}\binits{A.}}
(\byear{2014}).
\btitle{Understanding predictive information criteria for Bayesian models}.
\bjournal{Statistics and computing}
\bvolume{24}
\bpages{997--1016}.
\end{barticle}
\endbibitem

\bibitem[\protect\citeauthoryear{Glass, Guttmacher and
  Black}{2012}]{glass2012ending}
\begin{barticle}[author]
\bauthor{\bsnm{Glass},~\bfnm{Roger~I}\binits{R.~I.}},
  \bauthor{\bsnm{Guttmacher},~\bfnm{Alan~E}\binits{A.~E.}} \AND
  \bauthor{\bsnm{Black},~\bfnm{Robert~E}\binits{R.~E.}}
(\byear{2012}).
\btitle{Ending preventable child death in a generation}.
\bjournal{Journal of the American Medical Association}
\bvolume{308}
\bpages{141--142}.
\end{barticle}
\endbibitem

\bibitem[\protect\citeauthoryear{He et~al.}{2017}]{he2017national}
\begin{barticle}[author]
\bauthor{\bsnm{He},~\bfnm{Chunhua}\binits{C.}},
  \bauthor{\bsnm{Liu},~\bfnm{Li}\binits{L.}},
  \bauthor{\bsnm{Chu},~\bfnm{Yue}\binits{Y.}},
  \bauthor{\bsnm{Perin},~\bfnm{Jamie}\binits{J.}},
  \bauthor{\bsnm{Dai},~\bfnm{Li}\binits{L.}},
  \bauthor{\bsnm{Li},~\bfnm{Xiaohong}\binits{X.}},
  \bauthor{\bsnm{Miao},~\bfnm{Lei}\binits{L.}},
  \bauthor{\bsnm{Kang},~\bfnm{Leni}\binits{L.}},
  \bauthor{\bsnm{Li},~\bfnm{Qi}\binits{Q.}},
  \bauthor{\bsnm{Scherpbier},~\bfnm{Robert}\binits{R.}} \betal{et~al.}
(\byear{2017}).
\btitle{National and subnational all-cause and cause-specific child mortality
  in {C}hina, 1996--2015: a systematic analysis with implications for the
  {S}ustainable {D}evelopment {G}oals}.
\bjournal{The Lancet Global Health}
\bvolume{5}
\bpages{e186--e197}.
\end{barticle}
\endbibitem

\bibitem[\protect\citeauthoryear{Heaton et~al.}{2019}]{heaton2019case}
\begin{barticle}[author]
\bauthor{\bsnm{Heaton},~\bfnm{Matthew~J}\binits{M.~J.}},
  \bauthor{\bsnm{Datta},~\bfnm{Abhirup}\binits{A.}},
  \bauthor{\bsnm{Finley},~\bfnm{Andrew~O}\binits{A.~O.}},
  \bauthor{\bsnm{Furrer},~\bfnm{Reinhard}\binits{R.}},
  \bauthor{\bsnm{Guinness},~\bfnm{Joseph}\binits{J.}},
  \bauthor{\bsnm{Guhaniyogi},~\bfnm{Rajarshi}\binits{R.}},
  \bauthor{\bsnm{Gerber},~\bfnm{Florian}\binits{F.}},
  \bauthor{\bsnm{Gramacy},~\bfnm{Robert~B}\binits{R.~B.}},
  \bauthor{\bsnm{Hammerling},~\bfnm{Dorit}\binits{D.}},
  \bauthor{\bsnm{Katzfuss},~\bfnm{Matthias}\binits{M.}} \betal{et~al.}
(\byear{2019}).
\btitle{A case study competition among methods for analyzing large spatial
  data}.
\bjournal{Journal of Agricultural, Biological and Environmental Statistics}
\bvolume{24}
\bpages{398--425}.
\end{barticle}
\endbibitem

\bibitem[\protect\citeauthoryear{Held, Schr{\"o}dle and
  Rue}{2010}]{held2010posterior}
\begin{bincollection}[author]
\bauthor{\bsnm{Held},~\bfnm{Leonhard}\binits{L.}},
  \bauthor{\bsnm{Schr{\"o}dle},~\bfnm{Birgit}\binits{B.}} \AND
  \bauthor{\bsnm{Rue},~\bfnm{H{\aa}vard}\binits{H.}}
(\byear{2010}).
\btitle{Posterior and cross-validatory predictive checks: a comparison of MCMC
  and INLA}.
In \bbooktitle{Statistical Modelling and Regression Structures}
\bpages{91--110}.
\bpublisher{Springer}.
\end{bincollection}
\endbibitem

\bibitem[\protect\citeauthoryear{Holford}{1976}]{holford1976life}
\begin{barticle}[author]
\bauthor{\bsnm{Holford},~\bfnm{Theodore~R}\binits{T.~R.}}
(\byear{1976}).
\btitle{Life tables with concomitant information}.
\bjournal{Biometrics}
\bpages{587--597}.
\end{barticle}
\endbibitem

\bibitem[\protect\citeauthoryear{Holford}{1980}]{holford1980analysis}
\begin{barticle}[author]
\bauthor{\bsnm{Holford},~\bfnm{Theodore~R}\binits{T.~R.}}
(\byear{1980}).
\btitle{The analysis of rates and of survivorship using log-linear models}.
\bjournal{Biometrics}
\bpages{299--305}.
\end{barticle}
\endbibitem

\bibitem[\protect\citeauthoryear{Jha}{2012}]{jha2012counting}
\begin{barticle}[author]
\bauthor{\bsnm{Jha},~\bfnm{Prabhat}\binits{P.}}
(\byear{2012}).
\btitle{Counting the dead is one of the world's best investments to reduce
  premature mortality}.
\bjournal{Hypothesis}
\bvolume{10}
\bpages{e3}.
\end{barticle}
\endbibitem

\bibitem[\protect\citeauthoryear{Keenan et~al.}{2018}]{keenan2018azithromycin}
\begin{barticle}[author]
\bauthor{\bsnm{Keenan},~\bfnm{Jeremy~D}\binits{J.~D.}},
  \bauthor{\bsnm{Bailey},~\bfnm{Robin~L}\binits{R.~L.}},
  \bauthor{\bsnm{West},~\bfnm{Sheila~K}\binits{S.~K.}},
  \bauthor{\bsnm{Arzika},~\bfnm{Ahmed~M}\binits{A.~M.}},
  \bauthor{\bsnm{Hart},~\bfnm{John}\binits{J.}},
  \bauthor{\bsnm{Weaver},~\bfnm{Jerusha}\binits{J.}},
  \bauthor{\bsnm{Kalua},~\bfnm{Khumbo}\binits{K.}},
  \bauthor{\bsnm{Mrango},~\bfnm{Zakayo}\binits{Z.}},
  \bauthor{\bsnm{Ray},~\bfnm{Kathryn~J}\binits{K.~J.}},
  \bauthor{\bsnm{Cook},~\bfnm{Catherine}\binits{C.}} \betal{et~al.}
(\byear{2018}).
\btitle{Azithromycin to reduce childhood mortality in sub-Saharan Africa}.
\bjournal{New England Journal of Medicine}
\bvolume{378}
\bpages{1583--1592}.
\end{barticle}
\endbibitem

\bibitem[\protect\citeauthoryear{Keller et~al.}{2015}]{keller2015unified}
\begin{barticle}[author]
\bauthor{\bsnm{Keller},~\bfnm{Joshua~P}\binits{J.~P.}},
  \bauthor{\bsnm{Olives},~\bfnm{Casey}\binits{C.}},
  \bauthor{\bsnm{Kim},~\bfnm{Sun-Young}\binits{S.-Y.}},
  \bauthor{\bsnm{Sheppard},~\bfnm{Lianne}\binits{L.}},
  \bauthor{\bsnm{Sampson},~\bfnm{Paul~D}\binits{P.~D.}},
  \bauthor{\bsnm{Szpiro},~\bfnm{Adam~A}\binits{A.~A.}},
  \bauthor{\bsnm{Oron},~\bfnm{Assaf~P}\binits{A.~P.}},
  \bauthor{\bsnm{Lindstr{\"o}m},~\bfnm{Johan}\binits{J.}},
  \bauthor{\bsnm{Vedal},~\bfnm{Sverre}\binits{S.}} \AND
  \bauthor{\bsnm{Kaufman},~\bfnm{Joel~D}\binits{J.~D.}}
(\byear{2015}).
\btitle{A unified spatiotemporal modeling approach for predicting
  concentrations of multiple air pollutants in the multi-ethnic study of
  atherosclerosis and air pollution}.
\bjournal{Environmental health perspectives}
\bvolume{123}
\bpages{301--309}.
\end{barticle}
\endbibitem

\bibitem[\protect\citeauthoryear{Laird and Olivier}{1981}]{laird1981covariance}
\begin{barticle}[author]
\bauthor{\bsnm{Laird},~\bfnm{Nan}\binits{N.}} \AND
  \bauthor{\bsnm{Olivier},~\bfnm{Donald}\binits{D.}}
(\byear{1981}).
\btitle{Covariance analysis of censored survival data using log-linear analysis
  techniques}.
\bjournal{Journal of the American Statistical Association}
\bvolume{76}
\bpages{231--240}.
\end{barticle}
\endbibitem

\bibitem[\protect\citeauthoryear{Lee, Green and Ryan}{2017}]{lee2017poisson}
\begin{barticle}[author]
\bauthor{\bsnm{Lee},~\bfnm{Jarod~YL}\binits{J.~Y.}},
  \bauthor{\bsnm{Green},~\bfnm{Peter~J}\binits{P.~J.}} \AND
  \bauthor{\bsnm{Ryan},~\bfnm{Louise~M}\binits{L.~M.}}
(\byear{2017}).
\btitle{On the "Poisson Trick" and its Extensions for Fitting Multinomial
  Regression Models}.
\bjournal{arXiv preprint arXiv:1707.08538}.
\end{barticle}
\endbibitem

\bibitem[\protect\citeauthoryear{Li et~al.}{2019}]{li2019changes}
\begin{barticle}[author]
\bauthor{\bsnm{Li},~\bfnm{Zehang}\binits{Z.}},
  \bauthor{\bsnm{Hsiao},~\bfnm{Yuan}\binits{Y.}},
  \bauthor{\bsnm{Godwin},~\bfnm{Jessica}\binits{J.}},
  \bauthor{\bsnm{Martin},~\bfnm{Bryan~D}\binits{B.~D.}},
  \bauthor{\bsnm{Wakefield},~\bfnm{Jon}\binits{J.}},
  \bauthor{\bsnm{Clark},~\bfnm{Samuel~J}\binits{S.~J.}},
  \bauthor{\bparticle{with support from the United Nations Inter-agency
  Group~for} \bsnm{Child Mortality~Estimation}} \AND \bauthor{\bparticle{its
  technical~advisory} \bsnm{group}}
(\byear{2019}).
\btitle{Changes in the spatial distribution of the under-five mortality rate:
  Small-area analysis of 122 DHS surveys in 262 subregions of 35 countries in
  Africa}.
\bjournal{PloS one}
\bvolume{14}
\bpages{e0210645}.
\end{barticle}
\endbibitem

\bibitem[\protect\citeauthoryear{Lindgren, Rue and
  Lindstrom}{2011}]{lindgren2011explicit}
\begin{barticle}[author]
\bauthor{\bsnm{Lindgren},~\bfnm{Finn}\binits{F.}},
  \bauthor{\bsnm{Rue},~\bfnm{H{\aa}vard}\binits{H.}} \AND
  \bauthor{\bsnm{Lindstrom},~\bfnm{Johan}\binits{J.}}
(\byear{2011}).
\btitle{An explicit link between Gaussian fields and Gaussian Markov random
  fields: the stochastic partial differential equation approach}.
\bjournal{Journal of the Royal Statistical Society B}
\bvolume{73}
\bpages{423--498}.
\end{barticle}
\endbibitem

\bibitem[\protect\citeauthoryear{Liu et~al.}{2016a}]{liu2016integrated}
\begin{barticle}[author]
\bauthor{\bsnm{Liu},~\bfnm{Shiwei}\binits{S.}},
  \bauthor{\bsnm{Wu},~\bfnm{Xiaoling}\binits{X.}},
  \bauthor{\bsnm{Lopez},~\bfnm{Alan~D}\binits{A.~D.}},
  \bauthor{\bsnm{Wang},~\bfnm{Lijun}\binits{L.}},
  \bauthor{\bsnm{Cai},~\bfnm{Yue}\binits{Y.}},
  \bauthor{\bsnm{Page},~\bfnm{Andrew}\binits{A.}},
  \bauthor{\bsnm{Yin},~\bfnm{Peng}\binits{P.}},
  \bauthor{\bsnm{Liu},~\bfnm{Yunning}\binits{Y.}},
  \bauthor{\bsnm{Li},~\bfnm{Yichong}\binits{Y.}},
  \bauthor{\bsnm{Liu},~\bfnm{Jiangmei}\binits{J.}} \betal{et~al.}
(\byear{2016}a).
\btitle{An integrated national mortality surveillance system for death
  registration and mortality surveillance, China}.
\bjournal{Bulletin of the World Health Organization}
\bvolume{94}
\bpages{46--57}.
\end{barticle}
\endbibitem

\bibitem[\protect\citeauthoryear{Liu et~al.}{2016b}]{liu2016global}
\begin{barticle}[author]
\bauthor{\bsnm{Liu},~\bfnm{Li}\binits{L.}},
  \bauthor{\bsnm{Oza},~\bfnm{Shefali}\binits{S.}},
  \bauthor{\bsnm{Hogan},~\bfnm{Dan}\binits{D.}},
  \bauthor{\bsnm{Chu},~\bfnm{Yue}\binits{Y.}},
  \bauthor{\bsnm{Perin},~\bfnm{Jamie}\binits{J.}},
  \bauthor{\bsnm{Zhu},~\bfnm{Jun}\binits{J.}},
  \bauthor{\bsnm{Lawn},~\bfnm{Joy~E}\binits{J.~E.}},
  \bauthor{\bsnm{Cousens},~\bfnm{Simon}\binits{S.}},
  \bauthor{\bsnm{Mathers},~\bfnm{Colin}\binits{C.}} \AND
  \bauthor{\bsnm{Black},~\bfnm{Robert~E}\binits{R.~E.}}
(\byear{2016}b).
\btitle{Global, regional, and national causes of under-5 mortality in 2000--15:
  an updated systematic analysis with implications for the {S}ustainable
  {D}evelopment {G}oals}.
\bjournal{The Lancet}
\bvolume{388}
\bpages{3027--3035}.
\end{barticle}
\endbibitem

\bibitem[\protect\citeauthoryear{Mahapatra}{2010}]{mahapatra2010overview}
\begin{binproceedings}[author]
\bauthor{\bsnm{Mahapatra},~\bfnm{Prasanta}\binits{P.}}
(\byear{2010}).
\btitle{An overview of the sample registration system in India}.
In \bbooktitle{Prince Mahidol Award Conference \& Global Health Information
  Forum}
\bpages{27--30}.
\end{binproceedings}
\endbibitem

\bibitem[\protect\citeauthoryear{Maher et~al.}{2010}]{maher2010translating}
\begin{barticle}[author]
\bauthor{\bsnm{Maher},~\bfnm{D}\binits{D.}},
  \bauthor{\bsnm{Biraro},~\bfnm{S}\binits{S.}},
  \bauthor{\bsnm{Hosegood},~\bfnm{Victoria}\binits{V.}},
  \bauthor{\bsnm{Isingo},~\bfnm{R}\binits{R.}},
  \bauthor{\bsnm{Lutalo},~\bfnm{T}\binits{T.}},
  \bauthor{\bsnm{Mushati},~\bfnm{P}\binits{P.}},
  \bauthor{\bsnm{Ngwira},~\bfnm{B}\binits{B.}},
  \bauthor{\bsnm{Nyirenda},~\bfnm{M}\binits{M.}},
  \bauthor{\bsnm{Todd},~\bfnm{J}\binits{J.}},
  \bauthor{\bsnm{Zaba},~\bfnm{B}\binits{B.}} \betal{et~al.}
(\byear{2010}).
\btitle{Translating global health research aims into action: the example of the
  ALPHA network}.
\bjournal{Tropical Medicine \& International Health}
\bvolume{15}
\bpages{321--328}.
\end{barticle}
\endbibitem

\bibitem[\protect\citeauthoryear{McCormick
  et~al.}{2016}]{mccormick2016probabilistic}
\begin{barticle}[author]
\bauthor{\bsnm{McCormick},~\bfnm{Tyler~H}\binits{T.~H.}},
  \bauthor{\bsnm{Li},~\bfnm{Zehang~Richard}\binits{Z.~R.}},
  \bauthor{\bsnm{Calvert},~\bfnm{Clara}\binits{C.}},
  \bauthor{\bsnm{Crampin},~\bfnm{Amelia~C}\binits{A.~C.}},
  \bauthor{\bsnm{Kahn},~\bfnm{Kathleen}\binits{K.}} \AND
  \bauthor{\bsnm{Clark},~\bfnm{Samuel~J}\binits{S.~J.}}
(\byear{2016}).
\btitle{Probabilistic cause-of-death assignment using verbal autopsies}.
\bjournal{Journal of the American Statistical Association}
\bvolume{111}
\bpages{1036--1049}.
\end{barticle}
\endbibitem

\bibitem[\protect\citeauthoryear{Mikkelsen et~al.}{2015}]{mikkelsen2015global}
\begin{barticle}[author]
\bauthor{\bsnm{Mikkelsen},~\bfnm{Lene}\binits{L.}},
  \bauthor{\bsnm{Phillips},~\bfnm{David~E}\binits{D.~E.}},
  \bauthor{\bsnm{AbouZahr},~\bfnm{Carla}\binits{C.}},
  \bauthor{\bsnm{Setel},~\bfnm{Philip~W}\binits{P.~W.}},
  \bauthor{\bsnm{De~Savigny},~\bfnm{Don}\binits{D.}},
  \bauthor{\bsnm{Lozano},~\bfnm{Rafael}\binits{R.}} \AND
  \bauthor{\bsnm{Lopez},~\bfnm{Alan~D}\binits{A.~D.}}
(\byear{2015}).
\btitle{A global assessment of civil registration and vital statistics systems:
  monitoring data quality and progress}.
\bjournal{The Lancet}
\bvolume{386}
\bpages{1395--1406}.
\end{barticle}
\endbibitem

\bibitem[\protect\citeauthoryear{Murray et~al.}{2014}]{murray2014using}
\begin{barticle}[author]
\bauthor{\bsnm{Murray},~\bfnm{Christopher~JL}\binits{C.~J.}},
  \bauthor{\bsnm{Lozano},~\bfnm{Rafael}\binits{R.}},
  \bauthor{\bsnm{Flaxman},~\bfnm{Abraham~D}\binits{A.~D.}},
  \bauthor{\bsnm{Serina},~\bfnm{Peter}\binits{P.}},
  \bauthor{\bsnm{Phillips},~\bfnm{David}\binits{D.}},
  \bauthor{\bsnm{Stewart},~\bfnm{Andrea}\binits{A.}},
  \bauthor{\bsnm{James},~\bfnm{Spencer~L}\binits{S.~L.}},
  \bauthor{\bsnm{Vahdatpour},~\bfnm{Alireza}\binits{A.}},
  \bauthor{\bsnm{Atkinson},~\bfnm{Charles}\binits{C.}},
  \bauthor{\bsnm{Freeman},~\bfnm{Michael~K}\binits{M.~K.}} \betal{et~al.}
(\byear{2014}).
\btitle{Using verbal autopsy to measure causes of death: the comparative
  performance of existing methods}.
\bjournal{BMC {M}edicine}
\bvolume{12}
\bpages{5}.
\end{barticle}
\endbibitem

\bibitem[\protect\citeauthoryear{{United Nations}}{2015}]{un2015sdg}
\begin{bmisc}[author]
\bauthor{\bsnm{{United Nations}}}
(\byear{2015}).
\btitle{Transforming our world: the 2030 {A}genda for {S}ustainable
  {D}evelopment}.
\bhowpublished{Resolution adopted by the {G}eneral {A}ssembly on 25 {S}eptember
  2015}.
\end{bmisc}
\endbibitem

\bibitem[\protect\citeauthoryear{Nkengasong
  et~al.}{2020}]{nkengasong2020improving}
\begin{barticle}[author]
\bauthor{\bsnm{Nkengasong},~\bfnm{John}\binits{J.}},
  \bauthor{\bsnm{Gudo},~\bfnm{Eduardo}\binits{E.}},
  \bauthor{\bsnm{Macicame},~\bfnm{Ivalda}\binits{I.}},
  \bauthor{\bsnm{Maunze},~\bfnm{Xadreque}\binits{X.}},
  \bauthor{\bsnm{Amouzou},~\bfnm{Agbessi}\binits{A.}},
  \bauthor{\bsnm{Banke},~\bfnm{Kathryn}\binits{K.}},
  \bauthor{\bsnm{Dowell},~\bfnm{Scott}\binits{S.}} \AND
  \bauthor{\bsnm{Jani},~\bfnm{Ilesh}\binits{I.}}
(\byear{2020}).
\btitle{Improving birth and death data for African decision making}.
\bjournal{The Lancet Global Health}
\bvolume{8}
\bpages{e35--e36}.
\end{barticle}
\endbibitem

\bibitem[\protect\citeauthoryear{O'Brien et~al.}{2009}]{o2009burden}
\begin{barticle}[author]
\bauthor{\bsnm{O'Brien},~\bfnm{Katherine~L}\binits{K.~L.}},
  \bauthor{\bsnm{Wolfson},~\bfnm{Lara~J}\binits{L.~J.}},
  \bauthor{\bsnm{Watt},~\bfnm{James~P}\binits{J.~P.}},
  \bauthor{\bsnm{Henkle},~\bfnm{Emily}\binits{E.}},
  \bauthor{\bsnm{Deloria-Knoll},~\bfnm{Maria}\binits{M.}},
  \bauthor{\bsnm{McCall},~\bfnm{Natalie}\binits{N.}},
  \bauthor{\bsnm{Lee},~\bfnm{Ellen}\binits{E.}},
  \bauthor{\bsnm{Mulholland},~\bfnm{Kim}\binits{K.}},
  \bauthor{\bsnm{Levine},~\bfnm{Orin~S}\binits{O.~S.}},
  \bauthor{\bsnm{Cherian},~\bfnm{Thomas}\binits{T.}} \betal{et~al.}
(\byear{2009}).
\btitle{Burden of disease caused by Streptococcus pneumoniae in children
  younger than 5 years: global estimates}.
\bjournal{The Lancet}
\bvolume{374}
\bpages{893--902}.
\end{barticle}
\endbibitem

\bibitem[\protect\citeauthoryear{{WHO Collaborative Study Team on the Role of
  Breastfeeding on the Prevention of Infant Mortality}}{2001}]{who2001effect}
\begin{barticle}[author]
\bauthor{\bsnm{{WHO Collaborative Study Team on the Role of Breastfeeding on
  the Prevention of Infant Mortality}}}
(\byear{2001}).
\btitle{Effect of breastfeeding on infant and child mortality due to infectious
  diseases in less developed countries: a pooled analysis.}
\bjournal{The Lancet}
\bvolume{355}
\bpages{451--455}.
\end{barticle}
\endbibitem

\bibitem[\protect\citeauthoryear{Penny et~al.}{2016}]{penny2016public}
\begin{barticle}[author]
\bauthor{\bsnm{Penny},~\bfnm{Melissa~A}\binits{M.~A.}},
  \bauthor{\bsnm{Verity},~\bfnm{Robert}\binits{R.}},
  \bauthor{\bsnm{Bever},~\bfnm{Caitlin~A}\binits{C.~A.}},
  \bauthor{\bsnm{Sauboin},~\bfnm{Christophe}\binits{C.}},
  \bauthor{\bsnm{Galactionova},~\bfnm{Katya}\binits{K.}},
  \bauthor{\bsnm{Flasche},~\bfnm{Stefan}\binits{S.}},
  \bauthor{\bsnm{White},~\bfnm{Michael~T}\binits{M.~T.}},
  \bauthor{\bsnm{Wenger},~\bfnm{Edward~A}\binits{E.~A.}},
  \bauthor{\bparticle{Van~de} \bsnm{Velde},~\bfnm{Nicolas}\binits{N.}},
  \bauthor{\bsnm{Pemberton-Ross},~\bfnm{Peter}\binits{P.}} \betal{et~al.}
(\byear{2016}).
\btitle{Public health impact and cost-effectiveness of the {RTS,S/AS01} malaria
  vaccine: a systematic comparison of predictions from four mathematical
  models}.
\bjournal{The Lancet}
\bvolume{387}
\bpages{367--375}.
\end{barticle}
\endbibitem

\bibitem[\protect\citeauthoryear{Pfeffermann et~al.}{2013}]{pfeffermann2013new}
\begin{barticle}[author]
\bauthor{\bsnm{Pfeffermann},~\bfnm{Danny}\binits{D.}} \betal{et~al.}
(\byear{2013}).
\btitle{New important developments in small area estimation}.
\bjournal{Statistical Science}
\bvolume{28}
\bpages{40--68}.
\end{barticle}
\endbibitem

\bibitem[\protect\citeauthoryear{Phillips et~al.}{2015}]{phillips2015well}
\begin{barticle}[author]
\bauthor{\bsnm{Phillips},~\bfnm{David~E}\binits{D.~E.}},
  \bauthor{\bsnm{AbouZahr},~\bfnm{Carla}\binits{C.}},
  \bauthor{\bsnm{Lopez},~\bfnm{Alan~D}\binits{A.~D.}},
  \bauthor{\bsnm{Mikkelsen},~\bfnm{Lene}\binits{L.}},
  \bauthor{\bsnm{De~Savigny},~\bfnm{Don}\binits{D.}},
  \bauthor{\bsnm{Lozano},~\bfnm{Rafael}\binits{R.}},
  \bauthor{\bsnm{Wilmoth},~\bfnm{John}\binits{J.}} \AND
  \bauthor{\bsnm{Setel},~\bfnm{Philip~W}\binits{P.~W.}}
(\byear{2015}).
\btitle{Are well functioning civil registration and vital statistics systems
  associated with better health outcomes?}
\bjournal{The Lancet}
\bvolume{386}
\bpages{1386--1394}.
\end{barticle}
\endbibitem

\bibitem[\protect\citeauthoryear{Plummer}{2015}]{plummer2015cuts}
\begin{barticle}[author]
\bauthor{\bsnm{Plummer},~\bfnm{Martyn}\binits{M.}}
(\byear{2015}).
\btitle{Cuts in Bayesian graphical models}.
\bjournal{Statistics and Computing}
\bvolume{25}
\bpages{37--43}.
\end{barticle}
\endbibitem

\bibitem[\protect\citeauthoryear{Prentice et~al.}{1978}]{prentice1978analysis}
\begin{barticle}[author]
\bauthor{\bsnm{Prentice},~\bfnm{Ross~L}\binits{R.~L.}},
  \bauthor{\bsnm{Kalbfleisch},~\bfnm{John~D}\binits{J.~D.}},
  \bauthor{\bsnm{Peterson~Jr},~\bfnm{Arthur~V}\binits{A.~V.}},
  \bauthor{\bsnm{Flournoy},~\bfnm{Nancy}\binits{N.}},
  \bauthor{\bsnm{Farewell},~\bfnm{Vern~T}\binits{V.~T.}} \AND
  \bauthor{\bsnm{Breslow},~\bfnm{Norman~E}\binits{N.~E.}}
(\byear{1978}).
\btitle{The analysis of failure times in the presence of competing risks}.
\bjournal{Biometrics}
\bpages{541--554}.
\end{barticle}
\endbibitem

\bibitem[\protect\citeauthoryear{Rao et~al.}{2010}]{rao2010mortality}
\begin{barticle}[author]
\bauthor{\bsnm{Rao},~\bfnm{Chalapati}\binits{C.}},
  \bauthor{\bsnm{Soemantri},~\bfnm{Soeharsono}\binits{S.}},
  \bauthor{\bsnm{Djaja},~\bfnm{Sarimawar}\binits{S.}},
  \bauthor{\bsnm{Adair},~\bfnm{Timothy}\binits{T.}},
  \bauthor{\bsnm{Wiryawan},~\bfnm{Yuana}\binits{Y.}},
  \bauthor{\bsnm{Pangaribuan},~\bfnm{Lamria}\binits{L.}},
  \bauthor{\bsnm{Irianto},~\bfnm{Joko}\binits{J.}},
  \bauthor{\bsnm{Kosen},~\bfnm{Soewarta}\binits{S.}},
  \bauthor{\bsnm{Lopez},~\bfnm{Alan~D}\binits{A.~D.}} \betal{et~al.}
(\byear{2010}).
\btitle{Mortality in Central Java: results from the indonesian mortality
  registration system strengthening project}.
\bjournal{BMC Research Notes}
\bvolume{3}
\bpages{325}.
\end{barticle}
\endbibitem

\bibitem[\protect\citeauthoryear{Riebler et~al.}{2016}]{riebler2016intuitive}
\begin{barticle}[author]
\bauthor{\bsnm{Riebler},~\bfnm{Andrea}\binits{A.}},
  \bauthor{\bsnm{S{\o}rbye},~\bfnm{Sigrunn~H}\binits{S.~H.}},
  \bauthor{\bsnm{Simpson},~\bfnm{Daniel}\binits{D.}} \AND
  \bauthor{\bsnm{Rue},~\bfnm{H{\aa}vard}\binits{H.}}
(\byear{2016}).
\btitle{An intuitive Bayesian spatial model for disease mapping that accounts
  for scaling}.
\bjournal{Statistical Methods in Medical Research}
\bvolume{25}
\bpages{1145--1165}.
\end{barticle}
\endbibitem

\bibitem[\protect\citeauthoryear{Roberts et~al.}{2017}]{roberts2017cross}
\begin{barticle}[author]
\bauthor{\bsnm{Roberts},~\bfnm{David~R}\binits{D.~R.}},
  \bauthor{\bsnm{Bahn},~\bfnm{Volker}\binits{V.}},
  \bauthor{\bsnm{Ciuti},~\bfnm{Simone}\binits{S.}},
  \bauthor{\bsnm{Boyce},~\bfnm{Mark~S}\binits{M.~S.}},
  \bauthor{\bsnm{Elith},~\bfnm{Jane}\binits{J.}},
  \bauthor{\bsnm{Guillera-Arroita},~\bfnm{Gurutzeta}\binits{G.}},
  \bauthor{\bsnm{Hauenstein},~\bfnm{Severin}\binits{S.}},
  \bauthor{\bsnm{Lahoz-Monfort},~\bfnm{Jos{\'e}~J}\binits{J.~J.}},
  \bauthor{\bsnm{Schr{\"o}der},~\bfnm{Boris}\binits{B.}},
  \bauthor{\bsnm{Thuiller},~\bfnm{Wilfried}\binits{W.}} \betal{et~al.}
(\byear{2017}).
\btitle{Cross-validation strategies for data with temporal, spatial,
  hierarchical, or phylogenetic structure}.
\bjournal{Ecography}
\bvolume{40}
\bpages{913--929}.
\end{barticle}
\endbibitem

\bibitem[\protect\citeauthoryear{Rue and Held}{2005}]{rue2005gaussian}
\begin{bbook}[author]
\bauthor{\bsnm{Rue},~\bfnm{Havard}\binits{H.}} \AND
  \bauthor{\bsnm{Held},~\bfnm{Leonhard}\binits{L.}}
(\byear{2005}).
\btitle{Gaussian Markov random fields: theory and applications}.
\bpublisher{CRC Press}.
\end{bbook}
\endbibitem

\bibitem[\protect\citeauthoryear{Rue, Martino and
  Chopin}{2009}]{rue2009approximate}
\begin{barticle}[author]
\bauthor{\bsnm{Rue},~\bfnm{H{\aa}vard}\binits{H.}},
  \bauthor{\bsnm{Martino},~\bfnm{Sara}\binits{S.}} \AND
  \bauthor{\bsnm{Chopin},~\bfnm{Nicolas}\binits{N.}}
(\byear{2009}).
\btitle{Approximate Bayesian inference for latent Gaussian models by using
  integrated nested Laplace approximations}.
\bjournal{Journal of the Royal Statistical Society B}
\bvolume{71}
\bpages{319--392}.
\end{barticle}
\endbibitem

\bibitem[\protect\citeauthoryear{Sankoh and Byass}{2012}]{sankoh2012indepth}
\begin{bmisc}[author]
\bauthor{\bsnm{Sankoh},~\bfnm{Osman}\binits{O.}} \AND
  \bauthor{\bsnm{Byass},~\bfnm{Peter}\binits{P.}}
(\byear{2012}).
\btitle{The INDEPTH Network: filling vital gaps in global epidemiology}.
\end{bmisc}
\endbibitem

\bibitem[\protect\citeauthoryear{Simpson et~al.}{2017}]{simpson2017penalising}
\begin{barticle}[author]
\bauthor{\bsnm{Simpson},~\bfnm{Daniel}\binits{D.}},
  \bauthor{\bsnm{Rue},~\bfnm{H{\aa}vard}\binits{H.}},
  \bauthor{\bsnm{Riebler},~\bfnm{Andrea}\binits{A.}},
  \bauthor{\bsnm{Martins},~\bfnm{Thiago~G}\binits{T.~G.}},
  \bauthor{\bsnm{S{\o}rbye},~\bfnm{Sigrunn~H}\binits{S.~H.}} \betal{et~al.}
(\byear{2017}).
\btitle{Penalising model component complexity: A principled, practical approach
  to constructing priors}.
\bjournal{Statistical Science}
\bvolume{32}
\bpages{1--28}.
\end{barticle}
\endbibitem

\bibitem[\protect\citeauthoryear{Smith and Khaled}{2012}]{smith2012estimation}
\begin{barticle}[author]
\bauthor{\bsnm{Smith},~\bfnm{Michael~S}\binits{M.~S.}} \AND
  \bauthor{\bsnm{Khaled},~\bfnm{Mohamad~A}\binits{M.~A.}}
(\byear{2012}).
\btitle{Estimation of copula models with discrete margins via Bayesian data
  augmentation}.
\bjournal{Journal of the American Statistical Association}
\bvolume{107}
\bpages{290--303}.
\end{barticle}
\endbibitem

\bibitem[\protect\citeauthoryear{Snow et~al.}{1997}]{snow1997relation}
\begin{barticle}[author]
\bauthor{\bsnm{Snow},~\bfnm{Robert~W}\binits{R.~W.}},
  \bauthor{\bsnm{Omumbo},~\bfnm{Judy~A}\binits{J.~A.}},
  \bauthor{\bsnm{Lowe},~\bfnm{Brett}\binits{B.}},
  \bauthor{\bsnm{Molyneux},~\bfnm{Catherine~S}\binits{C.~S.}},
  \bauthor{\bsnm{Obiero},~\bfnm{Jacktone-O}\binits{J.-O.}},
  \bauthor{\bsnm{Palmer},~\bfnm{Ayo}\binits{A.}},
  \bauthor{\bsnm{Weber},~\bfnm{Martin~W}\binits{M.~W.}},
  \bauthor{\bsnm{Pinder},~\bfnm{Margaret}\binits{M.}},
  \bauthor{\bsnm{Nahlen},~\bfnm{Bernard}\binits{B.}},
  \bauthor{\bsnm{Obonyo},~\bfnm{Charles}\binits{C.}} \betal{et~al.}
(\byear{1997}).
\btitle{Relation between severe malaria morbidity in children and level of
  Plasmodium falciparum transmission in Africa}.
\bjournal{The Lancet}
\bvolume{349}
\bpages{1650--1654}.
\end{barticle}
\endbibitem

\bibitem[\protect\citeauthoryear{Soleman, Chandramohan and
  Shibuya}{2006}]{soleman2006verbal}
\begin{barticle}[author]
\bauthor{\bsnm{Soleman},~\bfnm{Nadia}\binits{N.}},
  \bauthor{\bsnm{Chandramohan},~\bfnm{Daniel}\binits{D.}} \AND
  \bauthor{\bsnm{Shibuya},~\bfnm{Kenji}\binits{K.}}
(\byear{2006}).
\btitle{Verbal autopsy: current practices and challenges}.
\bjournal{Bulletin of the World Health Organization}
\bvolume{84}
\bpages{239--245}.
\end{barticle}
\endbibitem

\bibitem[\protect\citeauthoryear{Speckman and Sun}{2003}]{speckman2003fully}
\begin{barticle}[author]
\bauthor{\bsnm{Speckman},~\bfnm{Paul~L}\binits{P.~L.}} \AND
  \bauthor{\bsnm{Sun},~\bfnm{Dongchu}\binits{D.}}
(\byear{2003}).
\btitle{Fully Bayesian spline smoothing and intrinsic autoregressive priors}.
\bjournal{Biometrika}
\bvolume{90}
\bpages{289--302}.
\end{barticle}
\endbibitem

\bibitem[\protect\citeauthoryear{{R Core Team}}{2013}]{team2013r}
\begin{barticle}[author]
\bauthor{\bsnm{{R Core Team}}}
(\byear{2013}).
\btitle{R: A language and environment for statistical computing}.
\bjournal{R {F}oundation for statistical computing}.
\end{barticle}
\endbibitem

\bibitem[\protect\citeauthoryear{Vos et~al.}{2020}]{vos2020global}
\begin{barticle}[author]
\bauthor{\bsnm{Vos},~\bfnm{Theo}\binits{T.}},
  \bauthor{\bsnm{Lim},~\bfnm{Stephen~S}\binits{S.~S.}},
  \bauthor{\bsnm{Abbafati},~\bfnm{Cristiana}\binits{C.}},
  \bauthor{\bsnm{Abbas},~\bfnm{Kaja~M}\binits{K.~M.}},
  \bauthor{\bsnm{Abbasi},~\bfnm{Mohammad}\binits{M.}},
  \bauthor{\bsnm{Abbasifard},~\bfnm{Mitra}\binits{M.}},
  \bauthor{\bsnm{Abbasi-Kangevari},~\bfnm{Mohsen}\binits{M.}},
  \bauthor{\bsnm{Abbastabar},~\bfnm{Hedayat}\binits{H.}},
  \bauthor{\bsnm{Abd-Allah},~\bfnm{Foad}\binits{F.}},
  \bauthor{\bsnm{Abdelalim},~\bfnm{Ahmed}\binits{A.}} \betal{et~al.}
(\byear{2020}).
\btitle{Global burden of 369 diseases and injuries in 204 countries and
  territories, 1990--2019: a systematic analysis for the Global Burden of
  Disease Study 2019}.
\bjournal{The Lancet}
\bvolume{396}
\bpages{1204--1222}.
\end{barticle}
\endbibitem

\bibitem[\protect\citeauthoryear{Wakefield
  et~al.}{2019}]{wakefield2019estimating}
\begin{barticle}[author]
\bauthor{\bsnm{Wakefield},~\bfnm{Jon}\binits{J.}},
  \bauthor{\bsnm{Fuglstad},~\bfnm{Geir-Arne}\binits{G.-A.}},
  \bauthor{\bsnm{Riebler},~\bfnm{Andrea}\binits{A.}},
  \bauthor{\bsnm{Godwin},~\bfnm{Jessica}\binits{J.}},
  \bauthor{\bsnm{Wilson},~\bfnm{Katie}\binits{K.}} \AND
  \bauthor{\bsnm{Clark},~\bfnm{Samuel~J}\binits{S.~J.}}
(\byear{2019}).
\btitle{Estimating under-five mortality in space and time in a developing world
  context}.
\bjournal{Statistical Methods in Medical Research}
\bvolume{28}
\bpages{2614--2634}.
\end{barticle}
\endbibitem

\bibitem[\protect\citeauthoryear{Walker, Bryce and
  Black}{2007}]{walker2007interpreting}
\begin{barticle}[author]
\bauthor{\bsnm{Walker},~\bfnm{Neff}\binits{N.}},
  \bauthor{\bsnm{Bryce},~\bfnm{Jennifer}\binits{J.}} \AND
  \bauthor{\bsnm{Black},~\bfnm{Robert~E}\binits{R.~E.}}
(\byear{2007}).
\btitle{Interpreting health statistics for policymaking: the story behind the
  headlines}.
\bjournal{The Lancet}
\bvolume{369}
\bpages{956--963}.
\end{barticle}
\endbibitem

\bibitem[\protect\citeauthoryear{Walker et~al.}{2013}]{walker2013global}
\begin{barticle}[author]
\bauthor{\bsnm{Walker},~\bfnm{Christa L~Fischer}\binits{C.~L.~F.}},
  \bauthor{\bsnm{Rudan},~\bfnm{Igor}\binits{I.}},
  \bauthor{\bsnm{Liu},~\bfnm{Li}\binits{L.}},
  \bauthor{\bsnm{Nair},~\bfnm{Harish}\binits{H.}},
  \bauthor{\bsnm{Theodoratou},~\bfnm{Evropi}\binits{E.}},
  \bauthor{\bsnm{Bhutta},~\bfnm{Zulfiqar~A}\binits{Z.~A.}},
  \bauthor{\bsnm{O'Brien},~\bfnm{Katherine~L}\binits{K.~L.}},
  \bauthor{\bsnm{Campbell},~\bfnm{Harry}\binits{H.}} \AND
  \bauthor{\bsnm{Black},~\bfnm{Robert~E}\binits{R.~E.}}
(\byear{2013}).
\btitle{Global burden of childhood pneumonia and diarrhoea}.
\bjournal{The Lancet}
\bvolume{381}
\bpages{1405--1416}.
\end{barticle}
\endbibitem

\bibitem[\protect\citeauthoryear{Wang et~al.}{2020}]{wang2020global}
\begin{barticle}[author]
\bauthor{\bsnm{Wang},~\bfnm{Haidong}\binits{H.}},
  \bauthor{\bsnm{Abbas},~\bfnm{Kaja~M}\binits{K.~M.}},
  \bauthor{\bsnm{Abbasifard},~\bfnm{Mitra}\binits{M.}},
  \bauthor{\bsnm{Abbasi-Kangevari},~\bfnm{Mohsen}\binits{M.}},
  \bauthor{\bsnm{Abbastabar},~\bfnm{Hedayat}\binits{H.}},
  \bauthor{\bsnm{Abd-Allah},~\bfnm{Foad}\binits{F.}},
  \bauthor{\bsnm{Abdelalim},~\bfnm{Ahmed}\binits{A.}},
  \bauthor{\bsnm{Abolhassani},~\bfnm{Hassan}\binits{H.}},
  \bauthor{\bsnm{Abreu},~\bfnm{Lucas~Guimar{\~a}es}\binits{L.~G.}},
  \bauthor{\bsnm{Abrigo},~\bfnm{Michael~RM}\binits{M.~R.}} \betal{et~al.}
(\byear{2020}).
\btitle{Global age-sex-specific fertility, mortality, healthy life expectancy
  (HALE), and population estimates in 204 countries and territories,
  1950--2019: a comprehensive demographic analysis for the Global Burden of
  Disease Study 2019}.
\bjournal{The Lancet}
\bvolume{396}
\bpages{1160--1203}.
\end{barticle}
\endbibitem

\bibitem[\protect\citeauthoryear{Wheldon
  et~al.}{2013}]{wheldon2013reconstructing}
\begin{barticle}[author]
\bauthor{\bsnm{Wheldon},~\bfnm{Mark~C}\binits{M.~C.}},
  \bauthor{\bsnm{Raftery},~\bfnm{Adrian~E}\binits{A.~E.}},
  \bauthor{\bsnm{Clark},~\bfnm{Samuel~J}\binits{S.~J.}} \AND
  \bauthor{\bsnm{Gerland},~\bfnm{Patrick}\binits{P.}}
(\byear{2013}).
\btitle{Reconstructing past populations with uncertainty from fragmentary
  data}.
\bjournal{Journal of the American Statistical Association}
\bvolume{108}
\bpages{96--110}.
\end{barticle}
\endbibitem

\bibitem[\protect\citeauthoryear{Yang et~al.}{2005}]{yang2005mortality}
\begin{barticle}[author]
\bauthor{\bsnm{Yang},~\bfnm{Gonghuan}\binits{G.}},
  \bauthor{\bsnm{Hu},~\bfnm{Jianping}\binits{J.}},
  \bauthor{\bsnm{Rao},~\bfnm{Ke~Quin}\binits{K.~Q.}},
  \bauthor{\bsnm{Ma},~\bfnm{Jeimin}\binits{J.}},
  \bauthor{\bsnm{Rao},~\bfnm{Chalapati}\binits{C.}} \AND
  \bauthor{\bsnm{Lopez},~\bfnm{Alan~D}\binits{A.~D.}}
(\byear{2005}).
\btitle{Mortality registration and surveillance in China: history, current
  situation and challenges}.
\bjournal{Population Health Metrics}
\bvolume{3}
\bpages{3}.
\end{barticle}
\endbibitem

\bibitem[\protect\citeauthoryear{You et~al.}{2015}]{you2015global}
\begin{barticle}[author]
\bauthor{\bsnm{You},~\bfnm{Danzhen}\binits{D.}},
  \bauthor{\bsnm{Hug},~\bfnm{Lucia}\binits{L.}},
  \bauthor{\bsnm{Ejdemyr},~\bfnm{Simon}\binits{S.}},
  \bauthor{\bsnm{Idele},~\bfnm{Priscila}\binits{P.}},
  \bauthor{\bsnm{Hogan},~\bfnm{Daniel}\binits{D.}},
  \bauthor{\bsnm{Mathers},~\bfnm{Colin}\binits{C.}},
  \bauthor{\bsnm{Gerland},~\bfnm{Patrick}\binits{P.}},
  \bauthor{\bsnm{New},~\bfnm{Jin~Rou}\binits{J.~R.}},
  \bauthor{\bsnm{Alkema},~\bfnm{Leontine}\binits{L.}} \betal{et~al.}
(\byear{2015}).
\btitle{Global, regional, and national levels and trends in under-5 mortality
  between 1990 and 2015, with scenario-based projections to 2030: a systematic
  analysis by the {UN} {I}nter-agency {G}roup for {C}hild {M}ortality
  {E}stimation}.
\bjournal{The Lancet}
\bvolume{386}
\bpages{2275--2286}.
\end{barticle}
\endbibitem

\end{thebibliography}

\end{document}